\documentclass[a4paper,usenatbib]{mnras}

\usepackage{amsmath}
\usepackage{amssymb}
\usepackage{newtxtext,newtxmath}
\usepackage[T1]{fontenc}
\usepackage{ae,aecompl}
\usepackage{graphicx}
\usepackage{times}
\title[The CSED in the EAGLE simulation]{The cosmic spectral energy distribution in the EAGLE simulation}

\author[M. Baes et al.]{%
Maarten Baes,$^1$
Ana Tr\v{c}ka,$^1$
Peter Camps,$^1$
Angelos Nersesian,$^{1,2,3}$
James Trayford,$^4$
\newauthor
Tom Theuns,$^5$
and
Wouter Dobbels$^1$
\\
$^1$Sterrenkundig Observatorium, Universiteit Gent, Krijgslaan 281 S9, 9000 Gent, Belgium \\
$^2$National Observatory of Athens, Institute for Astronomy, Astrophysics, Space Applications and Remote Sensing, Ioannou Metaxa and Vasileos Pavlou, \\
15236, Athens, Greece \\
$^3$Department of Astrophysics, Astronomy \& Mechanics, Faculty of Physics, University of Athens, Panepistimiopolis, GR 15784, Greece \\
$^4$Leiden Observatory, Leiden University, P.O. Box 9513, NL-2300 RA Leiden, The Netherlands \\
$^5$Institute for Computational Cosmology, Department of Physics, University of Durham, South Road, Durham DH1 3LE, UK
}

\date{Accepted 2019 January 22. Received 2019 January 14; in original form 2018 November 8.}

\pubyear{2019}

\begin{document}
\label{firstpage}
\pagerange{\pageref{firstpage}--\pageref{lastpage}}
\maketitle

\begin{abstract}
The cosmic spectral energy distribution (CSED) is the total emissivity as a function of wavelength of galaxies in a given cosmic volume. We compare the observed CSED from the UV to the submm to that computed from the EAGLE cosmological hydrodynamical simulation, post-processed with stellar population synthesis models and including dust radiative transfer using the SKIRT code. The agreement with the data is better than 0.15 dex over the entire wavelength range at redshift $z=0$, except at UV wavelengths where the EAGLE model overestimates the observed CSED by up to a factor 2. Global properties of the CSED as inferred from CIGALE fits, such as the stellar mass density, mean star formation density, and mean dust-to-stellar-mass ratio, agree to within better than 20 per cent. At higher redshift, EAGLE increasingly underestimates the CSED at optical-NIR wavelengths with the FIR/submm emissivity underestimated by more than a factor of 5 by redshift $z=1$. We believe that these differences are due to a combination of incompleteness of the EAGLE-SKIRT database, the small simulation volume and the consequent lack of luminous galaxies, and our lack of knowledge on the evolution of the characteristics of the interstellar dust in galaxies. The impressive agreement between the simulated and observed CSED at lower $z$ confirms that the combination of EAGLE and SKIRT dust processing yields a fairly realistic representation of the local Universe.
\end{abstract}


\maketitle

\newpage
\section{Introduction}

The cosmic spectral energy distribution (CSED) is a fundamental observational characteristics of the Universe. It represents the total electromagnetic power generated within a cosmological unit volume as a function of wavelength. In the wavelength range between the UV and the submm, the vast majority of the emission is emitted by stars, gas and dust within galaxies, and hence the CSED is a complex function of both the volume density of different galaxy types and the different processes that shape the SEDs of individual galaxies. 

Until a few years ago, most efforts to measure the CSED were concentrated on limited regions of the electromagnetic spectrum: UV \citep{2002AJ....124.1258W, 2005ApJ...619L..31B, 2005ApJ...619L..15W, 2011MNRAS.413.2570R}, optical \citep{2002MNRAS.336..907N, 2003ApJ...592..819B, 2009MNRAS.399.1106M}, near-infared \citep{2001MNRAS.326..255C, 2001ApJ...560..566K, 2009MNRAS.397..868S}, mid-infrared \citep{1990MNRAS.242..318S, 2006MNRAS.370.1159B} and far-infrared/submm \citep{2006A&A...448..525T, 2012MNRAS.421.3027B, 2016MNRAS.456.1999M}. The full CSED can then be obtained by combining these different measurements. However, this approach often results in mutually inconsistent results, due to different source selection criteria, different levels of completeness, photometric measurement discrepancies, or cosmic variance \citep{2002MNRAS.329..579C, 2010MNRAS.407.2131D, 2012MNRAS.427.3244D}.

The optimal approach to measure the CSED is to use a single and spectroscopically complete volume-limited sample with multi-wavelength data covering the entire UV-submm range. Such surveys have lately become available, with Galaxy And Mass Assembly \citep[GAMA:][]{2011MNRAS.413..971D, 2015MNRAS.452.2087L} probably being the most complete one in the local Universe. \citet{2012MNRAS.427.3244D} measured the local CSED from the UV to the NIR directly from GAMA data and used SED templates to extrapolate it to the submm range. \citet{2014MNRAS.439.1245K} considered the contribution of different galaxy types to the UV--NIR CSED and showed that all types contribute significantly to the ambient inter-galactic radiation field. \citet{2016MNRAS.455.3911D} used the GAMA Panchromatic Data Release to derive the first self-consistent measurement of the entire UV-submm CSED. Binning their galaxy sample in three redshift bins, they find a clear signature for evolution in the CSED over the past 2.3 Gyr. Very recently, \citet{2017MNRAS.470.1342A} took this study a significant step further. They included newly reduced panchromatic data from the Cosmological Origins Survey \citep[COSMOS:][]{2007ApJS..172....1S, 2015MNRAS.447.1014D, 2017MNRAS.464.1569A}, which enabled them to study the evolution of the CSED out to $z=1$. They demonstrated that the bolometric UV-submm energy output of the Universe has decreased by about a factor four over the past 8 Gyr. 

Obviously, the CSED is nothing but the joint contribution of the individual spectral energy distributions (SEDs) of all galaxies within a cosmological unit volume. Overall, the shape and normalisation of the CSED, and its evolution with redshift, are strong and purely observable constraints for models of galaxy formation and evolution \citep[e.g.,][]{2011MNRAS.410.2556D}. 

Galaxy formation and evolution models come in two broad classes: semi-analytical models and cosmological hydrodynamical simulations. The former have been around for more than two decades \citep[e.g.,][]{1993MNRAS.264..201K, 1994MNRAS.271..781C, 1999MNRAS.310.1087S}. Very recently, different versions of the GALFORM semi-analytical model \citep{2000MNRAS.319..168C, 2014MNRAS.439..264G, 2016MNRAS.462.3854L} were used to predict the CSED, generally showing good agreement between the models and the data at low redshifts \citep{2018MNRAS.474..898A, 2018arXiv180805208C}. 

Cosmological hydrodynamical simulations, on the other hand, have only fairly recently achieved sufficient realism and statistics to become a serious contender for the semi-analytical models. Modern cosmological hydrodynamical simulations such as Illustris \citep{2014MNRAS.444.1518V}, EAGLE \citep{2015MNRAS.446..521S}, MassiveBlack-II \citep{2015MNRAS.450.1349K}, MUFASA \citep{2016MNRAS.462.3265D}, and IllustrisTNG \citep{2018MNRAS.473.4077P} can reproduce many characteristics of the present-day galaxy population, including the stellar mass function, the size distribution, the bimodality in colours, and the supermassive black hole mass function. To the best of our knowledge, the CSED of cosmological hydrodynamical simulations has never been calculated in a self-consistent way over the entire UV-submm wavelength range. 

In order to do so, one needs to calculate the panchromatic SED of each galaxy in the simulation. This does not only include the stellar emission by different stellar populations, but also the distorting effect of interstellar dust. Indeed, dust attenuates roughly 30\% of all the starlight in normal star-forming galaxies, and re-emits it as thermal emission in the infrared \citep{1996A&A...306...61B, 2002MNRAS.335L..41P, 2016A&A...586A..13V}. This means that detailed 3D dust radiative transfer calculations are required. In the past few years, 3D dust radiative transfer has seen a remarkable development \citep{2013ARA&A..51...63S}, and such detailed self-consistent calculations in a realistic, galaxy-wide setting have become possible \citep{2010MNRAS.403...17J, 2011ApJ...743..159H, 2014A&A...571A..69D, 2014MNRAS.439.3868D, 2015MNRAS.449..243N, 2015A&A...576A..31S, 2015MNRAS.454.2381G, 2017MNRAS.469.3775G}. In particular, panchromatic dust radiative transfer post-processing can nowadays be applied to large numbers of simulated galaxies from cosmological hydrodynamical simulations \citep{2016MNRAS.462.1057C, 2018ApJS..234...20C, 2018arXiv180908239R}.

The goal of this paper is to compare the CSED of the EAGLE suite of simulations to the observed CSED as measured from GAMA observations. In Section~{\ref{EAGLE.sec}} we briefly describe the EAGLE simulations and the mock observations database we use for our study. In Section~{\ref{CSED0.sec}} we compare the EAGLE-SKIRT CSED of the local Universe to observational data from the GAMA survey, and we derive and discuss a number of important characteristics of the local Universe. In Section~{\ref{Diff.sec}} we compare the local Universe CSED corresponding to different EAGLE simulations, and discuss strong and weak convergence, and a variation of the subgrid model parameters. In Section~{\ref{CSEDz.sec}} we discuss the cosmic evolution of the EAGLE-SKIRT CSED out to $z\sim1$ and compare it again to observational data from GAMA and G10/COSMOS. In Section~{\ref{Summary.sec}} we summarise our results.

Throughout this paper, we adopt $H_0 = 67.77$~km~s$^{-1}$~Mpc$^{-1}$, the Planck cosmology value adopted by EAGLE \citep{2014A&A...571A..16P}.

\section{The simulations}
\label{EAGLE.sec}

\subsection{The EAGLE simulations}

EAGLE \citep[Evolution and Assembly of GaLaxies and their Environments:][]{2015MNRAS.446..521S} is a suite of cosmological hydrodynamics simulations. The simulations were run using an updated version of the $N$-body/SPH code GADGET-3 \citep{2005MNRAS.364.1105S}. Different runs correspond to different volumes (cubic volumes ranging from 25 to 100 comoving megaparsecs on a side), different resolutions, and different physical prescriptions for baryonic processes including star formation, AGN feedback and cooling. The main characteristics of the most important EAGLE runs are listed in Table~{\ref{EAGLEruns.tab}}.

\begin{table*}
\caption{Main characteristics of the different EAGLE runs used in this paper. Columns from left to right: EAGLE model name; comoving volume size; dark matter particle mass; initial baryonic particle mass; the total number of galaxies with $M_\star>10^{8.5}~M_\odot$ for all 29 snapshots combined; the fraction of this total number of galaxies with sufficient dust to be included in the EAGLE-SKIRT catalogue; notes about the subgrid physics recipes. The recalibrated high-resolution simulation, RecalL0025N0752, is the run on which the main analysis in this paper is based.}
\label{EAGLEruns.tab}
\centering
\begin{tabular}{cccccccl}
\hline\hline\\[-1mm]
EAGLE run & $L$ & $N_{\text{tot}}$ & $m_{\text{DM}}$ & $m_{\text{gas}}$ & $N_{\text{gal}}$ & $f_{\text{dusty}}$ & subgrid calibration \\
& [Mpc] & & [$M_\odot$] & [$M_\odot$] &  & [\%] & \\[2mm]
\hline\\[-1mm]
RefL0100N1504 & $100$ & $2\times1504^3$ & $9.70\times10^6$ & $1.81\times10^6$ & $371,728$ & $63.6$ & fiducial calibration \\
RefL0050N0752 & $50$ & $2\times752^3$ & $9.70\times10^6$ & $1.81\times10^6$ & $48,261$ & $65.1$ & fiducial calibration \\
AGNdT9L0050N0752 & $50$ & $2\times752^3$ & $9.70\times10^6$ & $1.81\times10^6$ & $48.278$ & $64.7$ & different AGN feedback \\
RefL0025N0376 & $25$ & $2\times376^3$ & $9.70\times10^6$ & $1.81\times10^6$ & $5742$ & $67.4$ & fiducial calibration \\
RefL0025N0752 & $25$ & $2\times752^3$ & $1.21\times10^6$ & $2.26\times10^5$ & $8279$ & $94.4$ & fiducial calibration \\
{\bf{RecalL0025N0752}} & $25$ & $2\times752^3$ & $1.21\times10^6$ & $2.26\times10^5$ & $5954$ & $95.7$ & recalibrated subgrid parameters \\[2mm]
\hline
\end{tabular}
\end{table*}

The EAGLE simulations track the cosmic evolution of dark matter, baryonic gas, stars and massive black holes. As most other cosmological simulations, EAGLE lacks the resolution and the detailed physical recipes to model the cold phase in the ISM. To prevent artificial fragmentation of star-forming gas, the EAGLE simulations impose a pressure floor, corresponding to a polytropic equation of state \citep{2008MNRAS.383.1210S}. As a consequence, the ISM does not consist of resolved molecular clouds, but rather of a fairly smoothly distributed, pressurised gas. The most important criterion for star formation in this simulated ISM is a metallicity-dependent density threshold \citep{2004ApJ...609..667S}. Star formation is implemented according to the observed Kennicutt-Schmidt law \citep{1959ApJ...129..243S, 1998ApJ...498..541K}, and with a \citet{2003PASP..115..763C} initial mass function. The stellar evolution and chemical enrichment prescriptions in EAGLE are taken from \citet{2009MNRAS.399..574W}.

The EAGLE simulations have been calibrated to reproduce the local Universe stellar mass function, the galaxy-central black hole mass relation, and the galaxy mass-size relation, as described by \citet{2015MNRAS.450.1937C}. The simulations were shown to be in reasonable to excellent agreement with many other observables not considered in the calibration, including the H$_2$ galaxy mass function \citep{2015MNRAS.452.3815L}, the relation between stellar mass and angular momentum \citep{2017MNRAS.464.3850L}, the supermassive black hole mass function \citep{2016MNRAS.462..190R},  the atomic gas properties of galaxies \citep{2017MNRAS.464.4204C}, the galaxy size evolution \citep{2017MNRAS.465..722F}, and the evolution of the star formation rate function \citep{2017MNRAS.472..919K}.

\subsection{The EAGLE-SKIRT database}

EAGLE doesn't include dust as a separate species. \citet{2016MNRAS.462.1057C} and \citet{2017MNRAS.470..771T} introduced an advanced framework to add interstellar dust to the EAGLE galaxies, and to calculate mock observables that fully take into account the absorption, scattering and thermal emission by this dust. The recipe includes a resampling procedure for star forming particles, the use of MAPPINGS SED templates \citep{2008ApJS..176..438G} to model dusty H{\sc{ii}} regions, and the inclusion of a diffuse dust distribution based on the distribution of metals in the gas phase. The final step in the procedure is a full 3D dust radiative transfer simulation using the SKIRT code. SKIRT \citep{2015A&C.....9...20C} is an open-source 3D Monte Carlo dust radiative transfer code, equipped with advanced grids for spatial discretisation \citep{2013A&A...554A..10S, 2014A&A...561A..77S}, a hybrid parallelisation scheme \citep{2017A&C....20...16V}, a library of input models \citep{2015A&C....12...33B}, and a suite of optimisation techniques \citep{2003MNRAS.343.1081B, 2011ApJS..196...22B, 2016A&A...590A..55B, 2013ARA&A..51...63S}.

\citet{2016MNRAS.462.1057C} used this framework to calculate mock infrared and submm observations for a limited set of EAGLE $z=0$ galaxies, selected to match a sample of nearby galaxies selected from the Herschel Reference Survey \citep[HRS,][]{2010PASP..122..261B}. The parameters in the post-processing scheme were calibrated to reproduce the observed HRS submm colours and dust scaling relations \citep{2012A&A...540A..54B, 2012A&A...540A..52C, 2014MNRAS.440..942C}. Subsequently, \citet{2017MNRAS.470..771T} used this calibrated recipe to calculate mock optical images, broadband fluxes, colours, and spectral indices for more than 30,000 local Universe EAGLE galaxies. One of their conclusions is that this radiative transfer recipe shows a marked improvement in the color versus stellar mass diagram over a simple dust-screen model. 

\citet{2018ApJS..234...20C} have used a refined version of this postprocessing recipe to populate a database of mock observations for the EAGLE simulations. The resulting EAGLE-SKIRT database contains mock UV to submm flux densities and rest-frame luminosities for nearly half a million simulated galaxies, distributed over 23 redshift slices from $z=0$ to $z=6$. The mock observations were calculated for six different EAGLE runs, corresponding to different box sizes, mass resolutions and physical ingredients (see Table~{\ref{EAGLEruns.tab}}). All these data are available in the public EAGLE database \citep{2016A&C....15...72M}. 

An important caveat is that only galaxies with at least 250 dust particles\footnote{The number of dust particles in a simulated EAGLE galaxy is defined as $N_{\text{dust}} = \max(N_{\text{coldgas}},N_{\text{SFR}})$, where $N_{\text{SFR}}$ and $N_{\text{coldgas}}$ indicate the number of (sub)particles in the sets representing the star-forming regions and cold gas particles, respectively. These sets may contain original SPH particles extracted from the EAGLE snapshot and/or resampled sub-particles replacing star-forming region candidates. See \citet[][\S3.1]{2018ApJS..234...20C} for details.} are considered in the EAGLE-SKIRT catalogue. As discussed in detail by \citet{2018ApJS..234...20C}, a minimum number of dust particles is required for the radiative transfer postprocessing to be meaningful, and 250 was found to be an appropriate number. This threshold leads to an incompleteness of the EAGLE-SKIRT catalogue, with a bias against red and dead early-type galaxies (with a high stellar mass, but little dust) and against low-mass galaxies (with low stellar masses {\em{and}} low dust masses). The level of incompleteness differs for the different EAGLE runs: in the high-resolution RecalL0025N0752 run, 95.7\% of all galaxies with stellar masses above $10^{8.5}~M_\odot$ are included in the database, whereas this fraction drops to 63.6\% for the largest-volume RefL0100N1504 run \citep[][Table~1]{2018ApJS..234...20C}.

\section{The local Universe CSED}
\label{CSED0.sec}

We calculated the CSED at $\langle z\rangle = 0.05$ as derived from the EAGLE-SKIRT database. We base our main analysis on the recalibrated 25 Mpc volume simulation, RecalL0025N0752, as this simulation has the highest resolution. An analysis of the effect of resolution, simulation volume and subgrid physics recipes will be presented in Section~{\ref{Diff.sec}}. Since the two last EAGLE snapshots correspond to $z = 0$ and $z = 0.1$, we averaged over these two snapshots. At each snapshot and in each broadband filter, we calculated the CSED by summing the observed flux densities of every single galaxy in the EAGLE-SKIRT database, and subsequently normalising the sum based on the snapshot co-moving volume and luminosity distance.  We calculated the CSED in 31 broadband filters covering the UV-submm wavelength regime.

\begin{figure*}
\includegraphics[width=0.94\textwidth]{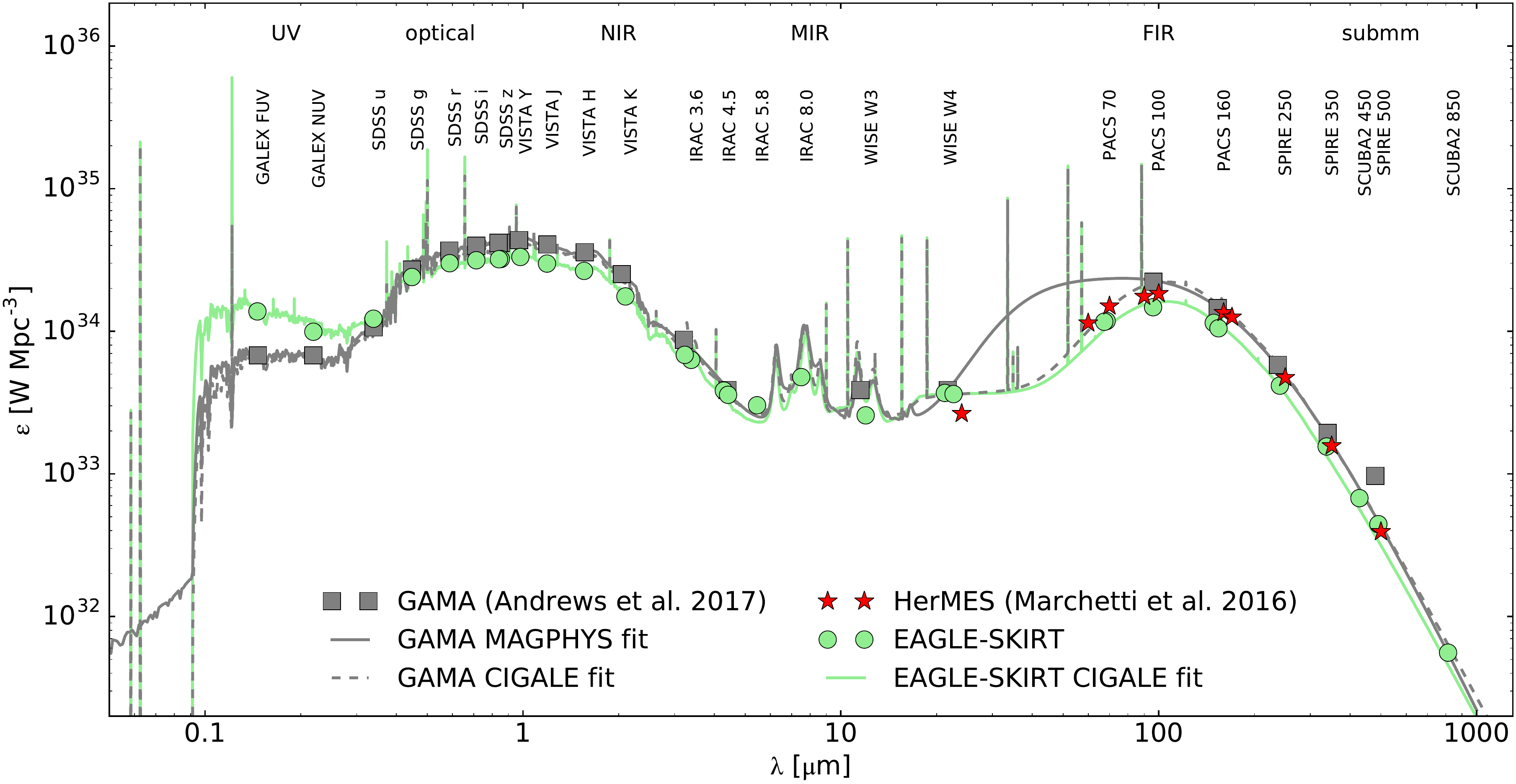}
\caption{The CSED in the local Universe ($\langle z\rangle = 0.05$). Solid grey squares are broadband observations from \citet{2017MNRAS.470.1342A}. The solid grey line is the best fitting MAGPHYS SED model as provided by \citet{2017MNRAS.470.1342A}, while the dashed grey line is the best fitting CIGALE SED model through the GAMA CSED broadband data points. The red stars are independent measurements of the infrared CSED based on the HerMES survey from \citet{2016MNRAS.456.1999M}. The green dots are the CSED corresponding to the EAGLE-SKIRT simulations, and the solid green line is the best fitting CIGALE fit to these data points. A number of broadband filters are indicated at the top.} 
\label{EAGLE-CSED0.fig}
\end{figure*}

In Figure~{\ref{EAGLE-CSED0.fig}} we compare the EAGLE-SKIRT CSED at $\langle z\rangle = 0.05$ to the observed GAMA CSED from \citet{2017MNRAS.470.1342A} for the redshift bin $0.02 < z < 0.08$, the lowest redshift bin they consider. The CSED is plotted as
\begin{equation}
\varepsilon \equiv \nu\epsilon_\nu \equiv \lambda\epsilon_\lambda
\end{equation}
and has units of total power per unit volume (W\,Mpc$^{-3}$). While the overall agreement between the GAMA observations and the EAGLE-SKIRT data points is very satisfactory, there are some minor differences to be noted, as we discuss next.

\subsection{The UV-optical-NIR CSED}
\label{UVOIR.sec}

First of all, the EAGLE-SKIRT results underestimate the GAMA CSED at red optical and near-infrared wavelengths ($\sim$0.6--4~$\mu$m). The difference is small (about 0.13 dex or about 30\%), but systematic. A first possible contributor to this systematic difference is the incompleteness of the EAGLE-SKIRT catalogue, in particular regarding the spheroidal galaxy population. \citet{2014MNRAS.439.1245K} find that more than half of the observed optical/NIR energy budget is dominated by galaxies with a significant spheroidal component. Herschel observations have shown that many early-types have dust masses below $5\times10^5~M_\odot$ \citep{2012ApJ...748..123S, 2013A&A...552A...8D}. Such galaxies, with a substantial contribution in the optical but almost no dust, could easily drop out of the EAGLE-SKIRT catalogue. 

Secondly, our simple method to estimate the CSED might also contribute to the systematic underestimation. We have estimated the CSED by simply adding the flux of each galaxy in the snapshot volume. As such, we might miss a non-negligible contribution from galaxies at the high-luminosity side of the luminosity function that is not properly sampled by the EAGLE RecalL0025N0752 simulation. \citet{2015MNRAS.452.2879T} have presented optical and NIR luminosity functions for the EAGLE simulations, based on mock fluxes that take dust absorption into account with a simple heuristic recipe. Their Figure~3 clearly demonstrates that the RecalL0025N0752 simulation misses the more luminous sources in the optical/NIR bands due to the relatively small simulation volume sampled. More specifically, the RecalL0025N0752 luminosity functions drop to zero around or even before the knee of the luminosity function. This insensitivity to the exponential cut-off at high luminosities, due to the small simulation volume, definitely contributes to the underestimation of the EAGLE-SKIRT CSED in the optical/NIR bands.

Finally, part of this discrepancy is due to the EAGLE simulation itself: \citet{2015MNRAS.452.2879T} note that their luminosity functions are consistent with a slight underestimate in the masses of more massive EAGLE galaxies, as already seen in the mass function shown by \citet{2015MNRAS.446..521S}.

In the SDSS g and u bands, the discrepancy between the EAGLE-SKIRT and the observed CSED is smaller than in the red and near-infrared filters, and at UV wavelengths, the EAGLE-SKIRT results even overestimate the observations (by 50\% in the GALEX NUV band, and even a factor 2 in the FUV band). The UV emission mainly originates from star forming regions, which are below the resolution limit for the EAGLE simulations. This resolution issue is handled using a subgrid approach, first employed by \citet{2010MNRAS.403...17J}: the star-forming regions are represented using template SEDs from the MAPPINGS library \citep{2008ApJS..176..438G}. These spherically symmetric models are controlled by different parameters, and it is well possible that the particular choice of these parameters can be further optimised. In particular, our calibration was based on submm colours and global dust scaling relations, and did not particularly focus on the UV wavelength regime \citep{2016MNRAS.462.1057C, 2018ApJS..234...20C}. In addition, the limited spatial resolution in the gas component in EAGLE results in a more homogeneous ISM distribution than we expect in actual galaxies, and EAGLE galaxies have systematically thicker discs, yielding a puffed up interstellar medium \citep{2017MNRAS.470..771T}. It is well-known that inhomogeneities and clumping can easily cause differences in the UV attenuation of an order of magnitude or more \citep[e.g.,][]{2000ApJ...528..799W, 2006ApJ...636..362I, 2015A&A...576A..31S}. 

In summary, we believe that the discrepancies between the GAMA and EAGLE-SKIRT CSED at UV to NIR wavelengths are primarily due to the incompleteness of the spheroidal galaxy population in the EAGLE-SKIRT catalogue, and an underestimation of the UV attenuation in the radiative transfer post-processing recipe.

\subsection{The infrared-submm CSED}
\label{IR.sec}

In the far-infrared region, the GAMA data points are again systematically higher than the EAGLE-SKIRT data points. At 100~$\mu$m, the difference is 0.18 dex, between 160 and 350~$\mu$m it is reduced to about 0.1 dex, but at 500 $\mu$m it increases again to 0.34 dex. However, \citet{2017MNRAS.470.1342A} indicate that their CSED is only poorly constrained or partially extrapolated due to lack of data beyond 24~$\mu$m. We therefore also show the independent measurements of the infrared CSED as obtained by \citet{2016MNRAS.456.1999M}, based on a very detailed analysis of Herschel and Spitzer data from the Herschel Multi-tiered Extragalactic Survey \citep[HerMES:][]{2012MNRAS.424.1614O}. Apart from the MIPS 24 $\mu$m band, where the agreement with GAMA was better, the EAGLE-SKIRT data agree much better with these independent measurements.  In particular at the longest wavelengths, nearly perfect agreement is found between EAGLE-SKIRT and the \citet{2016MNRAS.456.1999M} HerMES data. 

At far-infrared wavelength, i.e.\ between 60 and 250~$\mu$m, a small offset below 0.1 dex is observed. Given that the CSED at UV wavelengths overestimates the observations (Section~{\ref{UVOIR.sec}}), it seems likely that an underestimation of the UV attenuation in the radiative transfer post-processing recipe is the main reason for this offset. An additional factor can be the insensitivity of the RecalL0025N0752 simulation to luminous infrared sources (due to the limited simulation volume).

\subsection{Contribution of different populations}

\begin{figure*}
\includegraphics[width=0.94\textwidth]{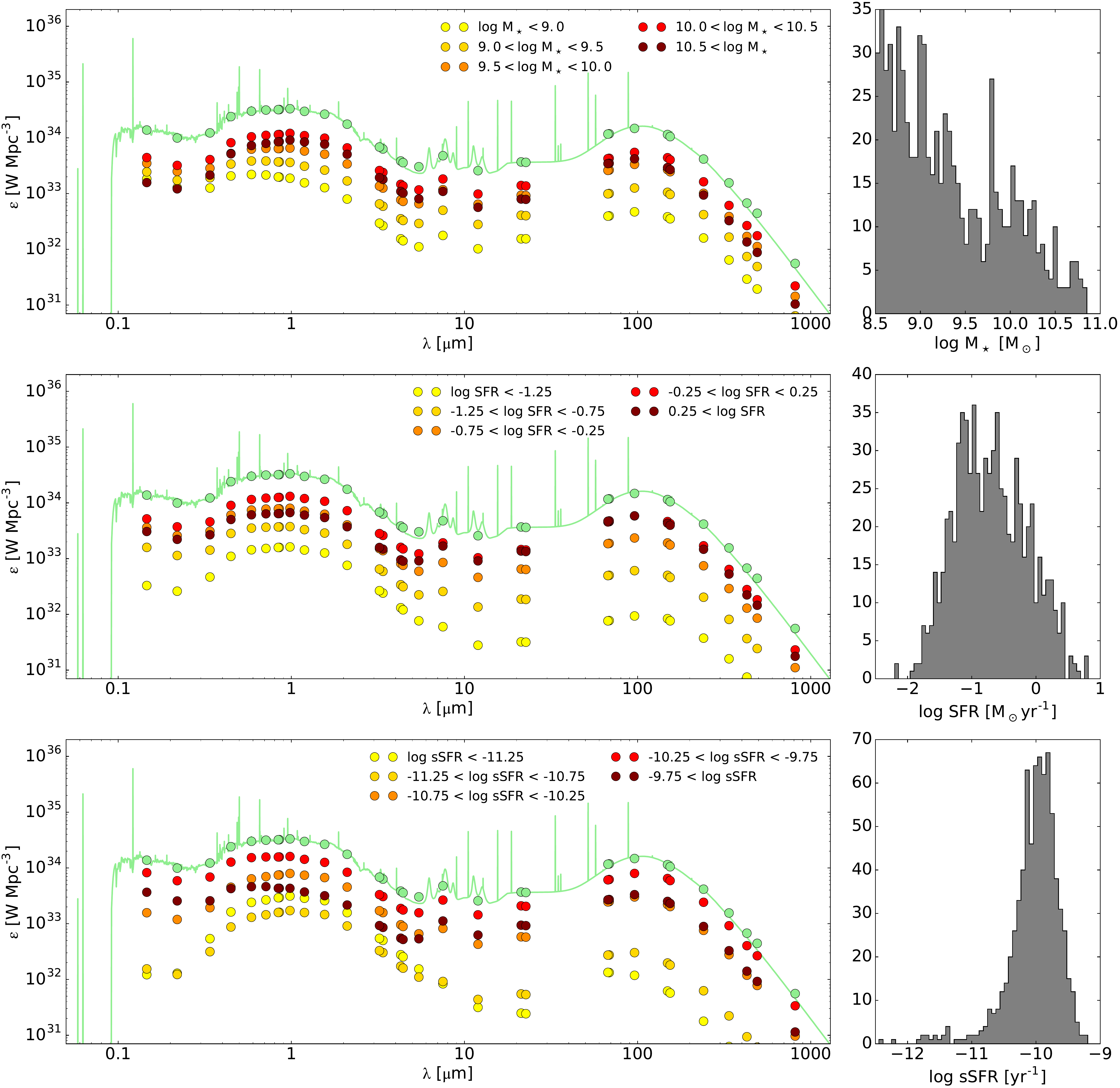}
\caption{The contributions of galaxies with different properties to the EAGLE-SKIRT CSED: stellar mass (top row), star formation rate (middle row), and specific star formation rate (bottom row). The right-hand panel on each row shows the histogram of the relevant quantity for the EAGLE-SKIRT sample. The left-hand panel shows the contributions of five different bins to the total EAGLE-SKIRT CSED.} 
\label{EAGLE-CSED-contributions.fig}
\end{figure*}

In Figure~{\ref{EAGLE-CSED-contributions.fig}} we split the local EAGLE-SKIRT CSED into contributions of galaxies in different bins in stellar mass (top row), star formation rate (middle row), and specific star formation rate (bottom row). 

The top row shows that  the CSED over the entire UV--submm wavelength range is dominated by galaxies within the stellar mass bin  with $10<\log(M_\star/M_\odot) < 10.5$). The more massive galaxy population with $\log(M_\star/M_\odot) > 10.5$ comes second in the optical--NIR range. Note that the galaxies in this most massive bin are underrepresented in the EAGLE-SKIRT database due to the threshold on the number of dust particles, and in general, are underrepresented in the RecalL0025N0752 simulation because of the small volume \citep{2015MNRAS.450.4486F, 2015MNRAS.452.2879T}. As expected, these most massive galaxies have a more modest contribution in the FIR-submm range, and a particularly low contribution in the UV. The population of increasingly lower-mass galaxies has an increasingly smaller contribution to the CSED, in spite of their increasing numbers. The lowest stellar mass bin has the smallest contribution along the entire spectrum.

The middle row shows that regular star-forming galaxies with a modest SFR around 1~$M_\odot$~yr$^{-1}$ contribute the bulk of the CSED over the entire wavelength range. Especially in the optical--NIR range they are clearly the most dominant contributor. The small population of galaxies in the highest SFR bin, with ${\text{SFR}} > 1.8~M_\odot\,{\text{yr}}^{-1}$ has an almost equal contribution in MIR--FIR range. The population of galaxies with ${\text{SFR}} < 0.06~M_\odot\,{\text{yr}}^{-1}$ has an almost negligible contribution to the CSED, even at optical--NIR wavelengths (but a fraction of these passive galaxies is missing in the EAGLE-SKIRT database).

The bottom row splits the contributions by populations in different sSFR bins. In this case, the contribution by galaxies centred around ${\text{sSFR}} \sim 10^{-10}~{\text{yr}}^{-1}$ is strongly dominant over the entire spectrum. This bin also corresponds to the largest number of galaxies. Interestingly, galaxies with a higher sSFR, although much less numerous, are the second-most important contributor in the UV and infrared range, while more passive galaxies with ${\text{sSFR}} \sim 10^{-10.5}~{\text{yr}}^{-1}$ contribute more to the optical--NIR CSED. Similarly, the galaxies with the lowest sSFR have the smallest contribution to the CSED in the UV and infrared range, whereas they contribute more in the optical--NIR range than the population of slightly higher sSFR galaxies. 

The bottom line is that main sequence galaxies with $M_\star\sim10^{10.25}~M_\odot$ and ${\text{sSFR}} \sim 10^{-10}~{\text{yr}}^{-1}$ dominate the local EAGLE-SKIRT CSED per dex in stellar mass or sSFR.

\subsection{Implications}

Overall, besides some minor differences that can readily be interpreted and explained, we find that the EAGLE-SKIRT CSED reproduces both the shape and the normalisation of the GAMA CSED very well over the entire UV-submm range. On the one hand, this is remarkable, given that no specific fitting or finetuning has been applied at this stage. This agreement can be interpreted as a confirmation that the combination of the EAGLE simulation and the SKIRT radiative transfer post-processing provides a solid representation of the present-day Universe. 

One factor that we have not yet taken into account is the sample variance. \citet{2017MNRAS.470.1342A} estimate the total uncertainty on their CSED to be of the order of 20 per cent, which is almost completely due to sample variance. We estimated the sample variance in our CSED estimate by splitting the 100 Mpc simulation volume into 64 individual 25 Mpc volumes, and calculating the variance based on the CSEDs of these 64 volumes. The level of sample variance obtained this way is around 65\% over the entire wavelength range, with only minor differences between the different wavelength regimes. This level of sample variance is consistent with the estimate of 56\% cosmic variance based on the empirical formulae from \citet{2010MNRAS.407.2131D} for a cubical 25 Mpc volume. Given these levels of uncertainty, the agreement between the GAMA and EAGLE-SKIRT CSEDs is impressive.

On the other hand, the close agreement in the local Universe might not be too surprising. The subgrid physics parameters in the EAGLE simulations were calibrated to reproduce a number of characteristics of the local Universe, including the local stellar mass function \citep{2015MNRAS.450.1937C, 2015MNRAS.446..521S}. In addition, the few free parameters in the SKIRT post-processing procedure were calibrated to reproduce the submm color-color relations and dust scaling relations as observed in the local Universe \citep{2016MNRAS.462.1057C}. The fact that this combination now also reproduces the UV-submm CSED to a large degree is a nice confirmation that these calibrations are fairly solid.

\subsection{Bolometric energy output of the local Universe}

From the estimates of the CSED in individual broadband filter bands, the logical next step is the determination of the bolometric energy output of the Universe\footnote{The quantity $\varepsilon_{\text{bol}}$ has the dimension energy per unit volume. However, it does not represent the bolometric radiative energy density, because it only contains the radiative energy {\em{generated}} in a representative unit volume, and not the energy connected to photons passing through this volume. This subtle distinction is important when considering the extragalactic background light \citep[e.g.,][]{2017MNRAS.470.1342A, 2018MNRAS.474..898A, 2018arXiv180805208C}.}, 
\begin{equation}
\varepsilon_{\text{bol}} = \int \varepsilon_\nu\,{\text{d}}\nu,
\end{equation}
where the integral covers the entire UV to submm wavelength range. This requires some interpolation scheme, and because of the complicated nature of translating broadband photometry into monochromatic flux densities, this integration is best done via SED fitting techniques \citep[for a discussion, see][]{2016AJ....152..102B}. 

The solid grey line in Figure~{\ref{EAGLE-CSED0.fig}} is a panchromatic energy balance template fit to the observed GAMA CSED data points, based on the MAGPHYS code \citep{2008MNRAS.388.1595D, 2018MNRAS.475.2891D}, and provided in tabular format by \citet{2017MNRAS.470.1342A}. The line fits all of the GAMA data points very well, except the SPIRE 500 $\mu$m data point that seems to be incompatible with the typical dust emission profile. From this fit, \citet{2017MNRAS.470.1342A} recover a value of $\varepsilon_{\text{bol}} = 1.26\times10^{35}$~W~Mpc$^{-3}$.

We repeated this analysis for our EAGLE-SKIRT CSED, but we used CIGALE, another popular SED fitting package \citep{2009A&A...507.1793N, 2018arXiv181103094B}. We used the CIGALE version 2018.0, equipped with the same model ingredients and parameter settings as \citet{Nersesian2019}. For the stellar populations, we have adopted the \citet{2003MNRAS.344.1000B} library of single stellar populations with a \citet{1955ApJ...121..161S} IMF, and a delayed and truncated star formation history model \citep{2016A&A...585A..43C}. Nebular emission was included, based on CLOUDY templates, and under the assumption that all the ionising radiation is absorbed by gas \citep{1998PASP..110..761F, 2011MNRAS.415.2920I}. For the dust, we adopt the THEMIS dust model \citep{2017A&A...602A..46J, 2017PASP..129d4102D}, and a modified version of the standard starburst-like attenuation \citep{2000ApJ...533..682C, 2016A&A...591A...6B}. In total, about $8\times10^7$ different models were considered in the fitting.

The solid green line in Figure~{\ref{EAGLE-CSED0.fig}} is the best fitting model. Integrating the CSED over the entire UV-submm wavelength range, we recover a value of $\varepsilon_{\text{bol}} = 9.94\times10^{34}$~W~Mpc$^{-3}$. This value is 0.10 dex below the GAMA value as measured by \citet{2017MNRAS.470.1342A}. This difference is mainly due to the difference in SED shape between the two models in the ill-covered region between 24 and 100 $\mu$m. The MAGPHYS fit to the GAMA CSED shows a remarkable ``boxy'' shape in this part of the spectrum. We believe that this shape is due to the setup of the MAGPHYS code, which models the dusty medium as a combination of four different components with adjustable temperatures \citep{2008MNRAS.388.1595D}. Such a behaviour has been noted in MAGPHYS SED fits to individual galaxies, where other panchromatic SED fitting codes do not show this feature \citep[e.g.,][]{2016A&A...589A..11P, 2018arXiv180904088H}.

The dashed grey line in Figure~{\ref{EAGLE-CSED0.fig}} is our own CIGALE fit to the GAMA CSED data points from \citet{2017MNRAS.470.1342A}. This SED model represents an equally good fit to the observed broadband data points, and avoids the boxy shape between 24 and 100 $\mu$m. Integrating the CSED now results in $\varepsilon_{\text{bol}} = 1.08\times10^{35}$~W~Mpc$^{-3}$, a difference of just 0.03 dex with our EAGLE-SKIRT value. To put these numbers in context: the bolometric energy output of the Universe corresponds to a single 50~W light bulb within a sphere of radius 1~AU. 

In conclusion, the total EAGLE-SKIRT energy output agrees very well with the GAMA observations, in spite of the differences in the CSED at UV and infrared wavelengths. This indicates that the underestimation of the EAGLE-SKIRT CSED at FIR wavelengths is mainly due to an underestimation of the UV attenuation, as suggested in Section~{\ref{IR.sec}}.

\subsection{Physical global characteristics of the local Universe}

\begin{table*}
\caption{Global characteristics of the local Universe as derived from SED model fits to the CSED. The third column gives the intrinsic SFR density and stellar mass contributing obtained by directly summing the SFRs and stellar masses of all galaxies in the EAGLE-SKIRT volume. The fourth column corresponds to the CIGALE fit to the EAGLE-SKIRT CSED as derived in this paper. The fifth column corresponds to our CIGALE fit to the observed GAMA CSED data points from \citet{2017MNRAS.470.1342A}. The last column corresponds to the detailed study by \citet{2018MNRAS.475.2891D}, based on MAGPHYS SED fits for individual galaxies from the GAMA, G10-COSMOS and 3D-HST surveys.}
\label{params.tab}
\centering
\begin{tabular}{cccccc}
\hline\hline\\[-1mm]
quantity 	& unit 	& EAGLE-SKIRT	& EAGLE-SKIRT	& GAMA		& GAMA		\\
		& 		& intrinsic		& CIGALE 	& CIGALE 	& MAGPHYS	\\[2mm]
\hline\\[-1mm]
$\log\varepsilon_{\text{bol}}$	& W~Mpc$^{-3}$ 	& $\cdots$ & $35.00$		& $35.03$		& $35.10$ \\[2mm]
\hline\\[-1mm]
$\rho_{\text{SFR}}$		& $M_\odot$~yr$^{-1}$~Mpc$^{-3}$	& $0.011$	& $0.017$ 		& $0.014$		& $0.011$ \\
$\Omega_{\star}$			&  $-$ & $0.0012$ 	& $0.0015$		& $0.0018$		& $0.0015$ \\
$\Omega_{\text{d}}$			&  $-$			& $\cdots$ & $6.5\times10^{-7}$		& $7.8\times10^{-7}$		& $1.1\times10^{-6}$ \\
$\Omega_{\text{d}}/\Omega_{\star}$ 	& $-$		& $\cdots$ & $4.4\times10^{-4}$		& $4.3\times10^{-4}$		& $7.4\times10^{-4}$ \\
$f_{\text{abs}}$ 			& $-$			& $\cdots$ & $0.27$		& $0.30$		& $\cdots$ \\
$A_{\text{FUV}}$ 			& mag 		& $\cdots$ & $0.91$		& $1.49$		& $1.52$ \\[2mm]
\hline
\end{tabular}
\end{table*}

The SED modelling exercise presented in the previous subsection was necessary to determine the total energy output of the Universe. But rather than just using it as a sophisticated integrator, we can use it to estimate a number of fundamental physical characteristics of the local Universe. Indeed, each SED model in the CIGALE library is characterised by a large number of physical parameters, such as the star formation rate, the stellar mass and dust mass. In fact, the goal of SED fitting usually is the determination and analysis of these parameters for a sample of galaxies \citep[e.g.][]{2012MNRAS.427..703S, 2015ApJS..219....8C, 2016A&A...591A...6B, 2016A&A...589A..11P, 2018MNRAS.475.2891D}. In a similar way, we can interpret the physical parameters of our best fitting SED model to the CSED data points to estimate the quantities as the cosmic star formation rate density, stellar mass density, and dust mass density of the Universe, and compare them to measurements directly based on the EAGLE particle data. 

This approach is not completely rigorous. In order to properly estimate, for example, the stellar mass density of the Universe, one should, for each individual galaxy in a representative cosmic volume, estimate the stellar mass using an SED fitting code, and subsequently sum all of these contributions. This will not necessarily yield the same answer as the stellar mass corresponding to the best fitting SED model to the sum of the individual galaxy SEDs. However, our CIGALE fit provides us with an easy and convenient shortcut to estimate these quantities, and it is instructive to compare these measurements to the actual values known in EAGLE. The most important characteristics are listed in Table~{\ref{params.tab}}. The third and fourth columns correspond to our CIGALE fits to the EAGLE-SKIRT and the GAMA CSED, respectively. The fifth column lists the values obtained by \citet{2018MNRAS.475.2891D}, based on MAGPHYS SED fits to individual galaxies from the combined GAMA, G10-COSMOS and 3D-HST surveys.

For the cosmic stellar mass density, we find a value $\rho_\star = 1.9\times10^8~M_\odot$~Mpc$^{-3}$, or equivalently $\Omega_\star = 0.0015$, in almost scarily good agreement with the results from \citet{2018MNRAS.475.2891D}. The value we obtain from our CIGALE fit to the GAMA CSED data points is 20\% higher. These numbers are also fully consistent with independent estimates of the local stellar mass density based on the same GAMA data \citep{2016MNRAS.462.4336M, 2017MNRAS.470..283W}, and fall nicely within the range of estimates based on independent studies \citep{2001MNRAS.326..255C, 2003ApJS..149..289B, 2004MNRAS.355..764P, 2005MNRAS.362.1233E, 2009MNRAS.398.2177L, 2013ApJ...767...50M}. It is also reassuring that our $\Omega_\star$ value determined from the CSED is completely consistent with the value obtained directly from the EAGLE stellar mass function: at $z\sim0.05$, \citet{2015MNRAS.450.4486F} found a value $\Omega_\star = 0.0016$. Given the completely different methodology, the different initial mass function\footnote{The EAGLE simulations use a \citet{2003PASP..115..763C} IMF, whereas a \citet{1955ApJ...121..161S} IMF is used for the CIGALE SED fitting.}, and the fact that the uncertainties on derived stellar masses due to stellar evolution models, even at a fixed IMF, are typically $\sim$0.3 dex \citep{2009ApJ...699..486C, 2012MNRAS.422.3285P}, this agreement is remarkable. Finally, when we estimate the "intrinsic" value for $\Omega_\star$ by simply summing the actual stellar masses of all galaxies in the 25 Mpc volume, we recover the value 0.0012, again in very good agreement.

For the cosmic star formation rate density, one of the most fundamental parameters in galaxy formation and evolution models, we find a value of 0.017~$M_\odot$\,yr$^{-1}$\,Mpc$^{-3}$. This value is 50\% higher than the GAMA value (0.011~$M_\odot$\,yr$^{-1}$\,Mpc$^{-3}$) reported by \citet{2018MNRAS.475.2891D}, which agrees perfectly with the intrinsic value obtained by summing the individual SFRs of all galaxies in the EAGLE 25 Mpc volume \citep{2017MNRAS.472..919K}. These values bracket the most credible recent literature values \citep{2011MNRAS.413.2570R, 2013MNRAS.433.2764G, 2014ARA&A..52..415M, 2016MNRAS.461..458D}. Given the lack of rigor in our method, and the intrinsic uncertainties in estimating SFRs from SED modelling due to the sensitive dependence on the assumed star formation history \citep{2018arXiv180904088H, 2018arXiv181103635C, 2018arXiv181103637L}, we find this agreement very satisfactory.

The most significant deviation between our results and the GAMA results corresponds to dust-related parameters. For the cosmic dust mass density we find a value $\rho_{\text{d}} = 8.3\times10^4~M_\odot~{\text{Mpc}}^{-3}$, or equivalently $\Omega_{\text{d}} = 6.5\times10^{-7}$. This is almost 75\% lower than the GAMA value of $1.1\times10^{-6}$ reported by \citet{2018MNRAS.475.2891D}, and at the bottom end of independent estimates for $\Omega_{\text{d}}$ in the local Universe, which range between $7\times10^{-7}$ and $2.7\times10^{-6}$ \citep{2011MNRAS.417.1510D, 2013MNRAS.433..695C, 2015MNRAS.452..397C, 2018MNRAS.479.1077B}. The low value we find is partly due to the underestimation of the UV attenuation and the resulting dearth of infrared emission, and to the incompleteness of our EAGLE-SKIRT catalogue. An additional factor is that the CIGALE fit to the CSED, on which our value for $\Omega_{\text{d}}$ is based, systematically lies below the EAGLE-SKIRT data points from 350 $\mu$m onwards (Figure~{\ref{EAGLE-CSED0.fig}}). This is a consequence of the infrared templates using in our CIGALE setup, which assume a power-law distribution in the radiation field strength \citep{2001ApJ...549..215D, 2011A&A...536A..88G}. This is suitable to fit the SEDs of individual galaxies, but apparently less ideal to fit the broader CSED that results from the sum of many individual SEDs with different cold dust temperatures.

The dust-to-stellar mass ratio $\Omega_{\text{d}}/\Omega_\star$ is a valuable tool to probe the evolution of galaxies, as it represents an observable measure of how much dust per unit stellar mass survives the various destruction processes in galaxies. Theoretical models outline the strong dependence of this quantity on the underlying star formation history \citep{2014A&A...562A..30S, 2017MNRAS.468.1505M, 2017MNRAS.465...54C}. For the dust-to-stellar-mass ratio of the local Universe we find $4.4\times10^{-4}$, in perfect agreement with the value obtained from our CIGALE fit to the GAMA data. The ratio based on the MAGPHYS analysis of GAMA galaxies from \citet{2018MNRAS.475.2891D} is, not surprisingly, almost 70\% higher. Observed dust-to-stellar mass ratios in individual galaxies vary over a wide range of values: the typical values for spiral galaxies are usually found in the range between $5\times10^{-4}$ and 0.01, whereas the values for early-type galaxies go down to $10^{-5}$ and below \citep{2011ApJ...738...89S,  2012ApJ...748..123S, 2012A&A...540A..52C, 2012MNRAS.419.3505D}.

The FUV attenuation is probably the most important difference between the EAGLE-SKIRT simulations and the GAMA observations. Both the CIGALE and MAGPHYS fits to the GAMA observations indicate $A_{\text{FUV}}\sim1.5$, whereas our EAGLE results yield an FUV attenuation below one magnitude. As discussed in Section~{\ref{CSED0.sec}}, the limited resolution of the EAGLE simulation, even in the higher-resolution 25 Mpc volume, combined with limitations in our EAGLE-SKIRT subgrid model for star-forming regions and the thickness of the ISM in the EAGLE subgrid physics, are probably largely responsible. In this context, it is important to mention that most SED fitting codes, including CIGALE and MAGPHYS, adopt a rather simple treatment of attenuation, which is essentially a one- or two-component absorbing screen model. Our SKIRT radiative transfer modelling of each EAGLE galaxy, on the other hand, computes the attenuation in a fully self-consistent way, including multiple anisotropic scattering and in a realistic 3D setting. On the level of individual galaxies, this can yield fairly different effective attenuation curves, as demonstrated using idealised models \citep{2000ApJ...528..799W, 2001MNRAS.326..733B, 2005MNRAS.359..171I}, and using EAGLE galaxies \citep{2015MNRAS.452.2879T, 2017MNRAS.470..771T}. Given that the stellar populations and ISM properties of galaxies vary systematically with mass, we believe that the FUV attenuation as derived from an SED fit to the global CSED should be taken with a grain of salt. A more detailed analysis of the FUV attenuation, based on SED fits to individual EAGLE galaxies, will be considered in future work.

An interesting quantity that can be calculated from the CIGALE fitting is the cosmic bolometric attenuation $f_{\text{abs}}$, defined as the fraction of the total energy output that is absorbed and re-emitted by dust. We find a value of 27\% for our EAGLE-SKIRT simulation, in very good agreement with the CIGALE fit to the GAMA data. These values are also very similar to the typical bolometric attenuation values found for individual star-forming galaxies of the local Universe. \citet{2016A&A...586A..13V} recently performed a detailed investigation of the bolometric attenuation of 239 late-type galaxies from the Herschel Reference Survey \citep{2010PASP..122..261B}, and found an average value of 32\%. They noted that this number is remarkably similar to the value obtained from previous studies, even including studies that hardly had any access to UV or submm data \citep{1991AJ....101..354S, 1995A&A...293L..65X, 2002MNRAS.335L..41P, 2011ApJ...738...89S}. \citet{2018A&A...620A.112B} present a larger study of the bolometric attenuation for more than 800 galaxies of different morphological types from the DustPedia sample \citep{2018A&A...609A..37C}. They find a slightly lower average value ($\langle f_{\text{abs}} \rangle = 19\%$), but note that their sample is missing high-luminosity and high-specific-star-formation-rate objects. When only considering late-type galaxies, they find a average value $\langle f_{\text{abs}} \rangle = 25\%$, very close to the mean value we find in our analysis.

\section{Comparison of different EAGLE simulations}
\label{Diff.sec}

\begin{figure*}
\includegraphics[width=0.98\textwidth]{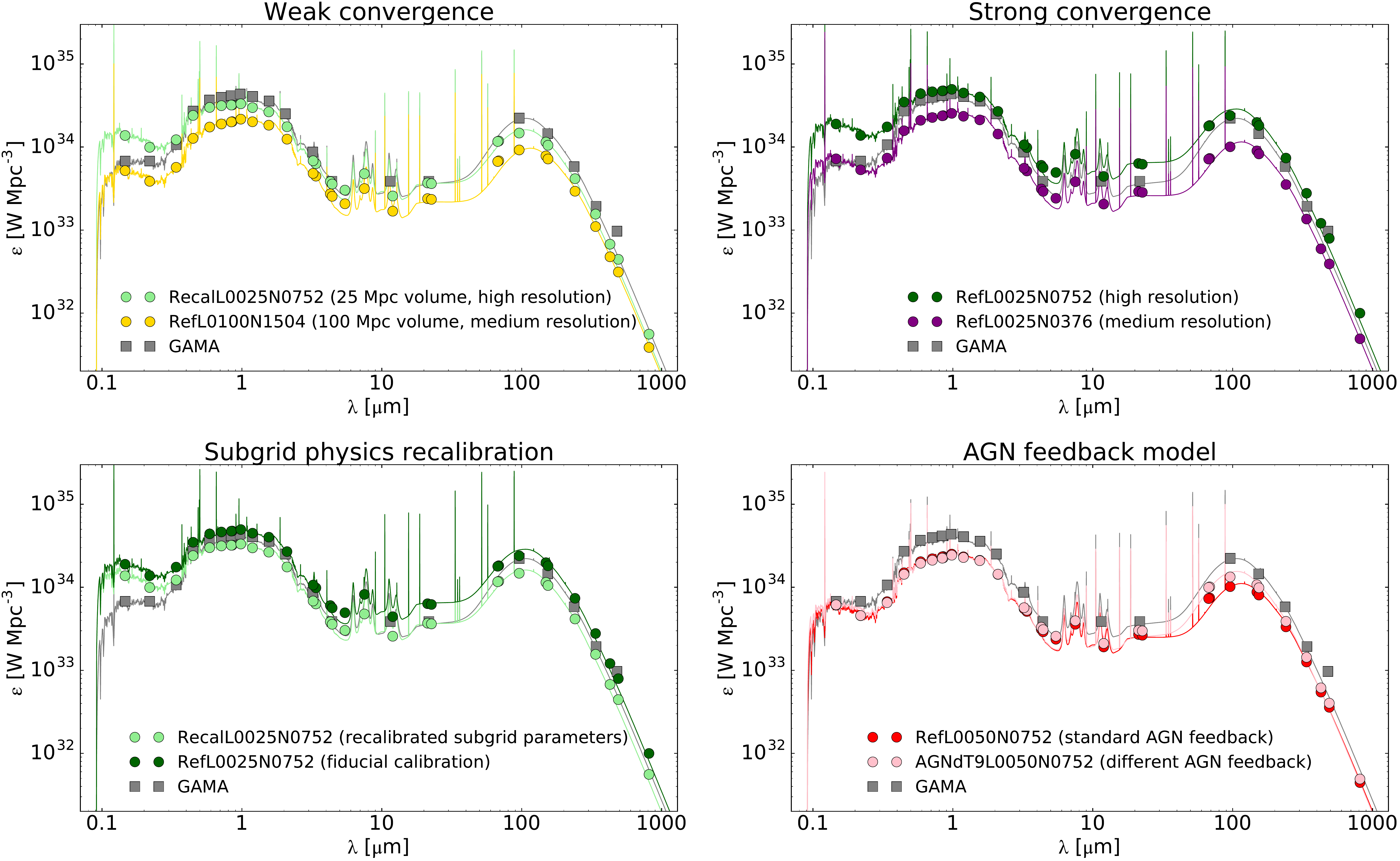}
\caption{A comparison of the CSED corresponding to different EAGLE runs. The models used are indicated in each panel. As in Figure~{\ref{EAGLE-CSED0.fig}}, the coloured dots are the broadband EAGLE-SKIRT CSED measurements, the solid grey squares are the broadband GAMA measurements from \citet{2017MNRAS.470.1342A}, and the solid lines represent CIGALE fits through the simulated/observed data. For details on the different simulations, see Table~{\ref{EAGLEruns.tab}}.} 
\label{EAGLE-CSED-modelvariation.fig}
\end{figure*}

\begin{figure*}
\includegraphics[width=0.98\textwidth]{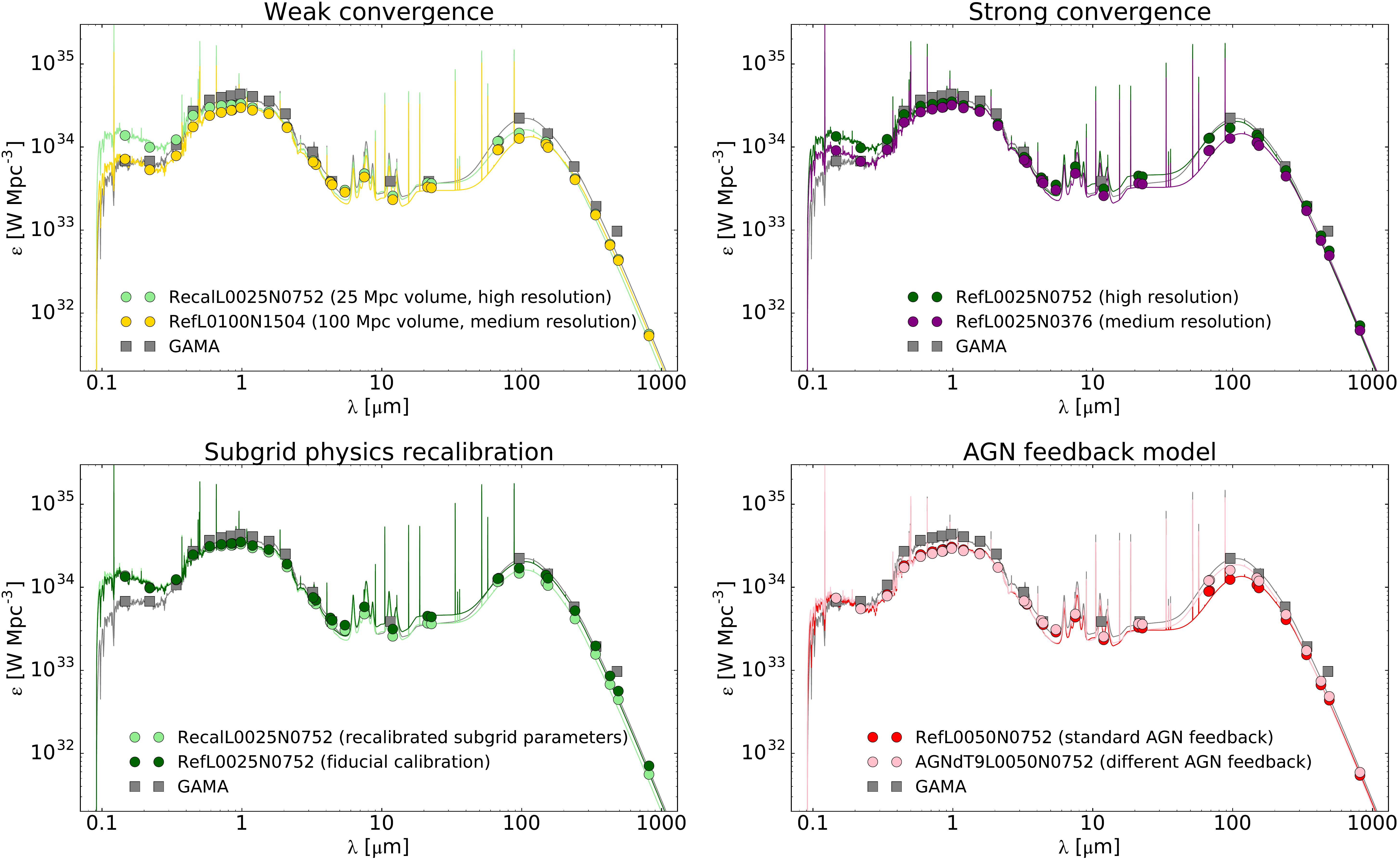}
\caption{A comparison of the normalised CSED corresponding to different EAGLE runs. This figure is identical to Figure~{\ref{EAGLE-CSED-modelvariation.fig}}, except that the CSED in each panel is normalised to the stellar mass density of the RecalL0025N0752 simulation.}
\label{EAGLE-CSED-modelvariation-normalised.fig}
\end{figure*}

All EAGLE-SKIRT results presented so far were based on the high-resolution RecalL0025N0752 simulation. In this section we compare these results to EAGLE-SKIRT CSEDs based on a number of other EAGLE simulations. In all cases, the CSEDs were calculated using exactly the same procedure; in particular, all galaxies from each run were post-processed using SKIRT with the same ``standard'' parameter values, as discussed by \citet{2018ApJS..234...20C}. 

\subsection{Strong and weak convergence}

The top-left panel of Figure~{\ref{EAGLE-CSED-modelvariation.fig}} compares the CSEDs of the RecalL0025N0752 and RefL0100N1504 simulations. The latter simulation is the reference EAGLE simulation, which has a volume 64 times larger, but a mass resolution eight times worse than the high-resolution RecalL0025N0752 run. Since the subgrid physics parameters in both simulations were independently calibrated to reproduce the galaxy properties in the local Universe, the comparison of both models can be regarded as a weak convergence test \citep{2015MNRAS.446..521S}. Such weak convergence tests have been done for various aspects of the EAGLE simulations, including the global stellar mass and SFR density \citep{2015MNRAS.450.4486F}, the relation between stellar mass and angular momentum \citep{2017MNRAS.464.3850L}, and optical luminosity functions \citep{2015MNRAS.452.2879T}.

It is immediately obvious that the RefL0100N1504 results systematically underestimate the RecalL0025N0752 results, and hence also the observed GAMA CSED. The difference between the two EAGLE CSEDs is on average about 0.2 dex, and reaches up to 0.4 dex at UV wavelengths. This can largely be explained by the incompleteness of the EAGLE-SKIRT database. For the RefL0100N1504 simulation, the threshold of 250 dust particles translates to a dust mass of about $4\times10^6~M_\odot$. Both low-mass late-type galaxies \citep{2010A&A...518L..52G, 2015A&A...574A.126G, 2011ApJ...738...89S} and massive elliptical galaxies \citep{2012ApJ...748..123S, 2013A&A...552A...8D} in the local Universe often have dust masses below this threshold. 

The top-left panel of Figure~{\ref{EAGLE-CSED-modelvariation-normalised.fig}} compares the same CSEDs as the top-left panel of Figure~{\ref{EAGLE-CSED-modelvariation.fig}}, but now the curves are normalised to the stellar mass density of the RecalL0025N0752 model. To some extent, this eliminates the differences in incompleteness of the EAGLE-SKIRT catalogues, and can highlight additional effects. Not surprisingly, the CSEDs for both models now agree nearly perfectly in the near-infrared. The largest differences between the normalised CSEDs are seen in the UV, with the RecalL0025N0752 model still 0.3 dex higher than the RefL0100N1504 model. This can be understood as the former model has a higher specific star formation rate density than the former, and thus a larger intrinsic UV output. In general, due to resolution effects, the RefL0100N1504 simulation is characterised by a relative overabundance of red/passive galaxies at $M_\star<10^9~M_\odot$ \citep{2015MNRAS.450.4486F, 2015MNRAS.452.2879T}. In spite of the higher specific star formation rate, the normalised FIR CSED of the RecalL0025N0752 model is not much higher than that of the RefL0100N1504 model. This might be due to the lower metallicity, and hence dust content, of RecalL0025N0752 galaxies: indeed, this run has a lower, and more realistic, mass-metallicity relation \citep{2015MNRAS.446..521S}.

The top-right panel of Figure~{\ref{EAGLE-CSED-modelvariation.fig}} can be regarded as a strong convergence test. This plot compares the CSEDs for models with exactly the same physical subgrid parameters (the parameters of the largest EAGLE volume) and the same simulation volume (25 Mpc on a side), but with a different resolution. The RefL0025N0376 simulation has the same resolution as the largest EAGLE volume simulation (RefL0100N1504), whereas the mass and spatial resolution of the RefL0025N0752 differ by factors of eight and two, respectively. It is no surprise that we see more or less the same effect as in the top-left panel: the higher threshold for the dust mass for the lower-resolution simulation causes an underestimation of the CSED over the entire wavelength range. Note, however, that this is most probably not the only reason: EAGLE simulations (and other cosmological hydro simulations) do not score well on strong convergence tests, which was exactly the reason why resolution-dependent recalibration has been implemented \citep{2015MNRAS.450.1937C, 2015MNRAS.446..521S}.

The top-right panel of Figure~{\ref{EAGLE-CSED-modelvariation-normalised.fig}} shows the corresponding version of this plot, but now again normalised to the stellar mass density of the RecalL0025N0752 model. A similar effect is noted as in the top-left panel: the higher resolution of the RefL0025N0752 simulation leads to a higher specific star formation rate, because feedback is less efficient. The result is a higher normalised CSED compared to the lower-resolution RefL0025N0376 simulation at UV and FIR wavelengths.

\subsection{Variation of the subgrid model parameters}

The bottom-left panel of Figure~{\ref{EAGLE-CSED-modelvariation.fig}} shows the effect of the resolution-dependent recalibration of the EAGLE subgrid physics parameters. This panel compares the CSEDs corresponding to two EAGLE simulations with exactly the same volume and resolution (the highest resolution of all EAGLE runs), but with different subgrid parameters. RefL0025L0752 uses the same subgrid physical parameters as the standard 100 Mpc simulation, whereas the RecalL0025L0752 simulation has different values for a number of subgrid physics parameters, including the characteristic density for star formation, and the temperature increase of the gas during AGN feedback \citep{2015MNRAS.446..521S}.

The effects of this recalibration are clearly visible: the RefL0025L0752 simulation systematically gives larger values for the CSED than the RecalL0025L0752 simulation, over the entire wavelength range. The difference between both CSEDs is roughly 0.2 dex, with the largest difference (0.25 dex) at submm wavelengths. This is no surprise, given that the intrinsic stellar mass density and the SFR density of the RefL0025L0752 simulation are respectively 0.15 dex and 0.20 dex higher than the corresponding values of the RecalL0025L0752 simulation \citep[see also][]{2015MNRAS.450.4486F}. When these CSEDs are normalised to the same stellar mass density (bottom-left panel of Figure~{\ref{EAGLE-CSED-modelvariation-normalised.fig}}), the differences become almost negligible. The normalised CSEDs of both models are almost identical over the entire UV--NIR wavelength range. The RecalL0025L0752 simulation has a slightly lower FIR-submm normalised CSED, because of the lower mean metallicity and hence dust content.

Finally, the bottom-right panel compares the CSEDs of two EAGLE runs with the same simulation volume and resolution, but with a different parameterisation of the AGN feedback. The RefL0050N0752 simulation uses the standard subgrid physics parameterisation, whereas the AGNdT9L0050N0752 model has adjusted AGN parameters in order to further improve the agreement with observations for high-mass galaxies. The main difference is an increased gas heating temperature increase for AGN feedback, which corresponds to more energetic but less frequent bursts, and more effective gas ejection. This improves the comparison to X-ray observations of the intracluster medium, at least on the scales of groups of galaxies \citep{2015MNRAS.446..521S}.

On the scales of individual galaxies, this change in the subgrid physics parameters does not have a strong effect. For example, the intrinsic stellar mass density of both models is identical, and the SFR density of the AGNdT9L0050N0752 model is only slightly higher (0.05 dex). As a result, both CSEDs are almost identical in the UV--NIR range, and the AGNdT9L0050N0752 has a slightly increased FIR CSED compared to the RefL0050N0752 CSED. The latter effect might be due to the fact that massive galaxies in the AGNdT9L0050N0752 run are slightly more compact \citep{2015MNRAS.446..521S}, and as a result, their dust will be slightly warmer on average. Due to the relatively poor resolution (a mass resolution eight times worse compared to the high-resolution simulation), the corresponding incompleteness of the EAGLE-SKIRT catalogue, and the lack of recalibration of these simulations, both CSEDs underestimate the observed GAMA CSED over the entire wavelength range.

\section{Cosmic evolution of the CSED}
\label{CSEDz.sec}

\begin{figure*}
\includegraphics[width=0.94\textwidth]{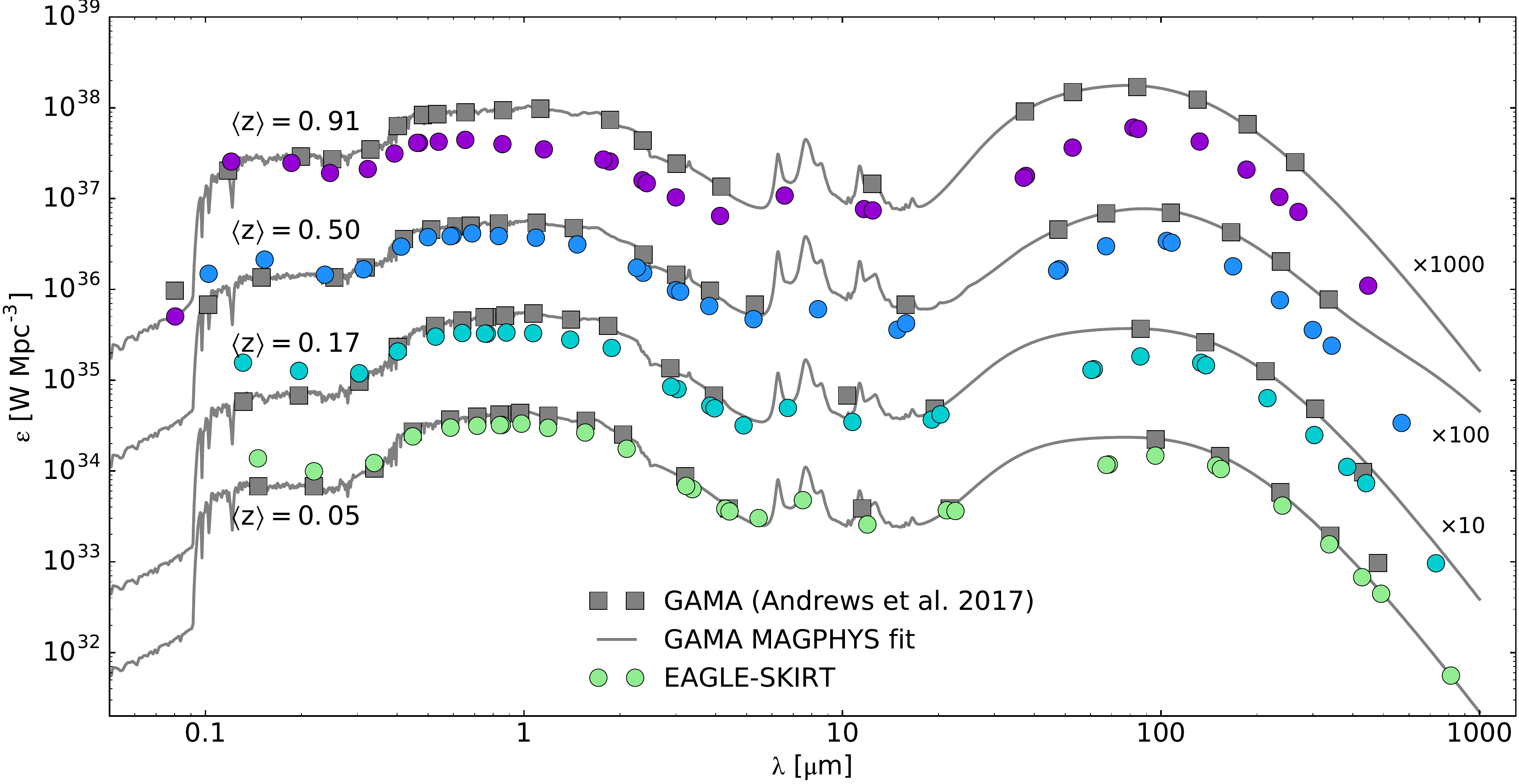}
\caption{The evolution of the CSED between $z=0$ and $z=1$. As in Figure~{\ref{EAGLE-CSED0.fig}}, the grey squares and solid grey lines are the GAMA broadband observations and best fitting MAGPHYS SED models from \citet{2017MNRAS.470.1342A}. The four different lines correspond to four different redshifts bins (as indicated on the left), and they have been shifted vertically for the sake of clarity. The coloured dots are the EAGLE-SKIRT CSED at the same mean redshifts.} 
\label{EAGLE-CSEDz.fig}
\end{figure*}

The comparison of the local Universe CSED provides a powerful test for galaxy evolution models, but an even more powerful test is the evolution of the CSED with cosmic time. Figure~{\ref{EAGLE-CSEDz.fig}} shows the CSED at four different redshifts between 0 and 1. As in Figure~{\ref{EAGLE-CSED0.fig}}, the grey squares correspond to the observed GAMA CSED from \citet{2017MNRAS.470.1342A}. For the redshift bins $0.02<z<0.08$ and $0.14<z<0.20$, the CSED is based on data from the standard equatorial GAMA fields \citep{2011MNRAS.413..971D, 2016MNRAS.455.3911D, 2015MNRAS.452.2087L}. For the redshift bins $0.45<z<0.56$ and $0.82<z<0.92$, the observed CSED is based on the G10/COSMOS data \citep{2015MNRAS.447.1014D, 2017MNRAS.464.1569A}. The solid grey lines are MAGPHYS fits to the observed CSED points, as provided by \citet{2017MNRAS.470.1342A}.

The coloured bullets correspond to the EAGLE-SKIRT simulation, derived in the same way as described in Section~{\ref{CSED0.sec}}. It is immediately obvious that the nice agreement between the EAGLE-SKIRT and GAMA results in the local Universe does not continue to higher redshifts. At $\langle z\rangle = 0.17$ and $\langle z\rangle = 0.50$ the underestimation of the CSED in the red and NIR domain is somewhat stronger than in the local Universe. The main issue, however, is the FIR-submm range, where the EAGLE-SKIRT results significantly underestimate the observed GAMA CSED. In the highest redshift bin corresponding to $\langle z\rangle = 0.91$, the disagreement is even worse. In the IRAC 1 band (corresponding to a rest-frame wavelength of $\sim2~\mu$m), the EAGLE-SKIRT estimate is a factor 5 lower than the observed data point, and the underestimation in the FIR/submm dust emission peak is even up to a factor 10. 

We note that a similar trend was also found by \citet{2018MNRAS.474..898A}: they find that the GALFORM semi-analytical model can reproduce the CSED for $z<0.3$ with a fairly good agreement, but the agreement becomes increasingly poor towards $z=1$.

The increasing disagreement between our EAGLE-SKIRT CSED and the GAMA observations with increasing redshift is most probably due to a combination of different factors. Firstly, there is the incompleteness of the EAGLE-SKIRT catalogue due to the threshold of at least 250 dust particles per galaxy, which becomes worse for higher $z$ because galaxies are less massive on average \citep{2018ApJS..234...20C}. A second aspect that comes into play, particularly at FIR/submm wavelengths, is that luminous infrared sources contribute increasingly more when moving to higher redshift. Different surveys have clearly shown that the infrared/submm luminosity function has a strong and rapid cosmic evolution out to $z\sim1$ \citep[e.g.][]{2009ApJ...707.1779E, 2010A&A...518L..23E, 2010A&A...518L..10D}. \citet{2016MNRAS.456.1999M} demonstrated that the characteristic luminosity in the SPIRE 250 $\mu$m band, $L_{250}^\star$, evolves as $(1+z)^{5.3}$. For the total infrared luminosity, they found an even stronger evolution, with $L_{\text{IR}}^\star \propto (1+z)^{6.0}$. In order to properly incorporate the contribution of rare but luminous galaxies to the CSED, the use of the small EAGLE-SKIRT simulation box is not ideal. 

It is, however, very unclear whether these two aspects can explain the differences, and other aspect might very well also contribute to, or even dominate, this disagreement. As already discussed, the subgrid physics in the EAGLE simulations were calibrated based on observed relations in the local Universe, and hence are not necessarily optimised for the Universe at higher redshift \citep{2015MNRAS.450.1937C, 2015MNRAS.446..521S}. We do note, however, that the EAGLE simulations show a reasonable level of agreement with the evolution of the galaxy stellar mass function \citep{2015MNRAS.450.4486F}, the mass-size relation \citep{2017MNRAS.465..722F}, and the cosmic star formation rate density \citep{2017MNRAS.472..919K}.

A similar argument applies to the EAGLE-SKIRT post-processing radiative transfer procedure. The calibration of this procedure was based on submm properties of galaxies in the local Universe \citep{2016MNRAS.462.1057C}, and turned out to be successful in also reproducing the optical colours in the local Universe \citep{2017MNRAS.470..771T}. However, as indicated by \citet{2018ApJS..234...20C}, this does not guarantee that these settings are also ideal for higher redshifts. In particular, for the construction of the EAGLE-SKIRT database, the same dust properties and fixed dust-to-metal ratio were adopted at all redshifts. It is uncertain, however, whether these assumptions are realistic, as different studies on this topic yield different conclusions. \citet{2013A&A...560A..26Z} performed a detailed study of the dust-to-metal ratio across a range of redshifts and sources, and do not find any obvious dependence of the dust-to-metals ratio on column density, galaxy type, redshift, or metallicity. Several other studies, including \citet{2013MNRAS.432.2112B} and \citet{2017MNRAS.471.1743D, DeVis2019}, do reveal a systematic evolution of the dust-to-metal ratio of galaxies. It is hence possible that our assumptions on the dust properties of the EAGLE galaxies are flawed, in particular towards higher redshift.

\citet{2011MNRAS.417.1510D} argued that the observed evolution of the dust mass function is difficult to explain using standard dust evolution models and requires a combination of various processes, including a possible evolution of dust grain properties. \citet{2018MNRAS.479.1077B} compares the local dust mass function with the predictions of the semi-analytical model of \citet{2017MNRAS.471.3152P} and the cosmological hydrodynamical model of \citet{2016MNRAS.457.3775M, 2017MNRAS.468.1505M}, both of which incorporate recipes for dust production and destruction. Both sets of theoretical predictions have difficulties in reproducing the dust mass function at both the low and high dust mass regimes, indicating that fundamental properties as stardust condensation efficiencies and dust grain growth time-scales are poorly understood. \citet{2018MNRAS.479.1077B} conclude that the current theoretical models for dust evolution need to be improved. We agree that more work is required in this field, and we particularly hope that a more integrated approach combining cosmological hydrodynamical simulations and dust evolution modelling \citep{2015MNRAS.449.1625B, 2016ApJ...831..147Z, 2016MNRAS.457.3775M, 2017MNRAS.468.1505M, 2018MNRAS.478.2851M, 2018MNRAS.478.4905A} will move this field forward.

\section{Summary and outlook}
\label{Summary.sec}

We have presented, for the first time, the local Universe cosmic spectral energy distribution (CSED) as derived from a cosmological hydrodynamical simulation (EAGLE RecalL0025L0752), combined with a dust radiative transfer post-processing algorithm (SKIRT). The CSED was obtained by simply summing the flux densities of individual galaxies in the EAGLE-SKIRT database recently presented by \citet{2018ApJS..234...20C}. Overall, we find an excellent agreement between the EAGLE-SKIRT CSED and the observed CSED based on GAMA observations, with some relatively minor differences:
\begin{itemize}
\item In the UV regime, the EAGLE-SKIRT CSED overestimates the observations, or conversely, underestimates the attenuation. This is likely a result of the limited resolution of the EAGLE simulations and the limitations in the SKIRT subgrid treatment of star forming regions. 
\item At optical and near-infrared wavelengths, the EAGLE-SKIRT CSED slightly underestimates the GAMA observations by about 0.13 dex. This is due to a combination of the incompleteness of the EAGLE-SKIRT database, which excludes galaxies with dust masses below $\sim5\times10^5~M_\odot$, the small volume of the EAGLE-SKIRT simulation and the resulting insensitivity for luminous galaxies, and a small EAGLE underestimation of the stellar mass function for massive galaxies. 
\item In the far-infrared and submm regime, our EAGLE-SKIRT results significantly underestimate the GAMA data points, which are only poorly constrained. This discrepancy largely disappears when our results are compared to independent estimates of the FIR/submm CSED based on deep Spitzer and Herschel data. The remaining underestimation of $\sim0.1$ dex is probably mainly due to the underestimation of the attenuation at UV wavelengths. 
\item Splitting the local EAGLE-SKIRT CSED into different stellar mass, SFR and sSFR populations, we find that main sequence galaxies with $M_\star\sim10^{10.25}~M_\odot$ and ${\text{sSFR}} \sim 10^{-10}~{\text{yr}}^{-1}$ are the dominant contributors to the local CSED over the entire UV--submm wavelength range.
\end{itemize}
Based on a physically motivated SED fit with the CIGALE code, we derive a number of global characteristics of the local Universe.
\begin{itemize}
\item Our EAGLE-SKIRT estimate of the bolometric energy output of the local Universe is in excellent agreement with the value obtained from GAMA observations. It is equivalent to a single 50~W light bulb in a sphere with radius 1 AU.
\item The cosmic star formation rate density derived from our CIGALE fit to the EAGLE-SKIRT CSED is about 50\% higher than the GAMA value, but perfectly within the range of independent values quoted in the recent literature for $z\sim0.05$. 
\item For the cosmic dust density and the cosmic dust-to-stellar-mass ratio, we find values that are some 70\% lower than the GAMA based estimates. This is due to a combination of the underestimation of the UV attenuation, the incompleteness of the EAGLE-SKIRT database, and the difficulties CIGALE has to properly fit the CSED data points at the longest wavelengths.
\item We obtain the result that 27\% of all radiative energy emitted by stars in the local Universe is absorbed by dust and re-emitted as thermal radiation in the infrared regime. This number is in very good agreement with the typical numbers found for the bolometric attenuation in individual galaxies. 
\end{itemize}
Putting this all together, we interpret this excellent agreement between the simulated and observed local CSED as a confirmation that the combination of the EAGLE simulation and the SKIRT radiative transfer post-processing provides a reliable mock representation of the present-day Universe. 

Unfortunately, the nice agreement that we found in the local Universe does not hold at higher redshifts. Even at $\langle z\rangle = 0.17$, we find significant deviations between the EAGLE-SKIRT results and the GAMA results, and the magnitude of these differences only increases out to $z\sim1$. The most important deviations are an increasing underestimation of the optical-NIR CSED, and an even greater discrepancy at far-infrared wavelengths. We believe that this discrepancy can be attributed to a combination of factors, including the incompleteness of the EAGLE-SKIRT database, the limited volume of the EAGLE 25 Mpc simulation (and hence the poor coverage of the high-luminosity end of the luminosity function), and our lack of knowledge on the evolution of the characteristics of the interstellar dust in galaxies. Indeed, one important caveat in our methodology is the assumption that the dust properties and the dust-to-metal ratio do not vary with increasing redshifts, but this assumption might be far too simplistic.

As the CSED is nothing but the joint contribution of the individual spectral energy distributions of all galaxies within a cosmological unit volume, it is a strong constraint for models of galaxy formation and evolution. We have shown in this paper that the EAGLE cosmological simulation, when combined with realistic mock observations based on detailed radiative transfer modelling, successfully withstands the comparison to the observed CSED. On the other hand, the CSED is not the ultimate test. A more stringent test would be to compare the EAGLE simulation to the CSED split up in different luminosity bins, i.e.\ luminosity functions covering the entire UV to submm wavelength range. Ideally, the luminosity functions could take into account information of EAGLE simulations at different resolutions, where the higher-resolution simulations constrain the low-luminosity regime, and the large volume simulations the high-luminosity tail. This is beyond the scope of this paper, and will be considered for future work.

\section*{Acknowledgements}

The authors thank the anonymous referee for constructive comments that improved this paper.  M.B., A.T., A.N., and W.D.\ gratefully acknowledge support from the Flemish Fund for Scientific Research (FWO-Vlaanderen), and from DustPedia, a European Union's Seventh Framework Programme (FP7) for research, technological development and demonstration under grant agreement no. 606874.

The EAGLE-SKIRT database on which this work was based used the DiRAC Data Centric system at Durham University, operated by the Institute for Computational Cosmology on behalf of the STFC DiRAC HPC Facility (http://www.dirac.ac.uk). This equipment was funded by BIS National E-infrastructure capital grant ST/K00042X/1, STFC capital grants ST/H008519/1 and ST/K00087X/1, STFC DiRAC Operations grant ST/K003267/ 1, and Durham University. DiRAC is part of the National E-Infrastructure.

\bibliographystyle{mnras}
\bibliography{EAGLE}

\begin{thebibliography}{}
\makeatletter
\relax
\def\mn@urlcharsother{\let\do\@makeother \do\$\do\&\do\#\do\^\do\_\do\%\do\~}
\def\mn@doi{\begingroup\mn@urlcharsother \@ifnextchar [ {\mn@doi@}
  {\mn@doi@[]}}
\def\mn@doi@[#1]#2{\def\@tempa{#1}\ifx\@tempa\@empty \href
  {http://dx.doi.org/#2} {doi:#2}\else \href {http://dx.doi.org/#2} {#1}\fi
  \endgroup}
\def\mn@eprint#1#2{\mn@eprint@#1:#2::\@nil}
\def\mn@eprint@arXiv#1{\href {http://arxiv.org/abs/#1} {{\tt arXiv:#1}}}
\def\mn@eprint@dblp#1{\href {http://dblp.uni-trier.de/rec/bibtex/#1.xml}
  {dblp:#1}}
\def\mn@eprint@#1:#2:#3:#4\@nil{\def\@tempa {#1}\def\@tempb {#2}\def\@tempc
  {#3}\ifx \@tempc \@empty \let \@tempc \@tempb \let \@tempb \@tempa \fi \ifx
  \@tempb \@empty \def\@tempb {arXiv}\fi \@ifundefined
  {mn@eprint@\@tempb}{\@tempb:\@tempc}{\expandafter \expandafter \csname
  mn@eprint@\@tempb\endcsname \expandafter{\@tempc}}}

\bibitem[\protect\citeauthoryear{{Andrews}, {Driver}, {Davies}, {Kafle},
  {Robotham}  \& {Wright}}{{Andrews} et~al.}{2017a}]{2017MNRAS.464.1569A}
{Andrews} S.~K.,  {Driver} S.~P.,  {Davies} L.~J.~M.,  {Kafle} P.~R.,
  {Robotham} A.~S.~G.,   {Wright} A.~H.,  2017a, \mn@doi [\mnras]
  {10.1093/mnras/stw2395}, \href
  {http://adsabs.harvard.edu/abs/2017MNRAS.464.1569A} {464, 1569}

\bibitem[\protect\citeauthoryear{{Andrews} et~al.,}{{Andrews}
  et~al.}{2017b}]{2017MNRAS.470.1342A}
{Andrews} S.~K.,  et~al., 2017b, \mn@doi [\mnras] {10.1093/mnras/stx1279},
  \href {http://adsabs.harvard.edu/abs/2017MNRAS.470.1342A} {470, 1342}

\bibitem[\protect\citeauthoryear{{Andrews}, {Driver}, {Davies}, {Lagos}  \&
  {Robotham}}{{Andrews} et~al.}{2018}]{2018MNRAS.474..898A}
{Andrews} S.~K.,  {Driver} S.~P.,  {Davies} L.~J.~M.,  {Lagos} C.~d.~P.,
  {Robotham} A.~S.~G.,  2018, \mn@doi [\mnras] {10.1093/mnras/stx2843}, \href
  {http://adsabs.harvard.edu/abs/2018MNRAS.474..898A} {474, 898}

\bibitem[\protect\citeauthoryear{{Aoyama}, {Hou}, {Hirashita}, {Nagamine}  \&
  {Shimizu}}{{Aoyama} et~al.}{2018}]{2018MNRAS.478.4905A}
{Aoyama} S.,  {Hou} K.-C.,  {Hirashita} H.,  {Nagamine} K.,   {Shimizu} I.,
  2018, \mn@doi [\mnras] {10.1093/mnras/sty1431}, \href
  {http://adsabs.harvard.edu/abs/2018MNRAS.478.4905A} {478, 4905}

\bibitem[\protect\citeauthoryear{{Babbedge} et~al.,}{{Babbedge}
  et~al.}{2006}]{2006MNRAS.370.1159B}
{Babbedge} T.~S.~R.,  et~al., 2006, \mn@doi [\mnras]
  {10.1111/j.1365-2966.2006.10547.x}, \href
  {http://adsabs.harvard.edu/abs/2006MNRAS.370.1159B} {370, 1159}

\bibitem[\protect\citeauthoryear{{Baes} \& {Camps}}{{Baes} \&
  {Camps}}{2015}]{2015A&C....12...33B}
{Baes} M.,  {Camps} P.,  2015, \mn@doi [Astronomy and Computing]
  {10.1016/j.ascom.2015.05.006}, \href
  {http://adsabs.harvard.edu/abs/2015A%26C....12...33B} {12, 33}

\bibitem[\protect\citeauthoryear{{Baes} \& {Dejonghe}}{{Baes} \&
  {Dejonghe}}{2001}]{2001MNRAS.326..733B}
{Baes} M.,  {Dejonghe} H.,  2001, \mn@doi [\mnras]
  {10.1046/j.1365-8711.2001.04626.x}, \href
  {http://adsabs.harvard.edu/abs/2001MNRAS.326..733B} {326, 733}

\bibitem[\protect\citeauthoryear{{Baes} et~al.,}{{Baes}
  et~al.}{2003}]{2003MNRAS.343.1081B}
{Baes} M.,  et~al., 2003, \mn@doi [\mnras] {10.1046/j.1365-8711.2003.06770.x},
  \href {http://adsabs.harvard.edu/abs/2003MNRAS.343.1081B} {343, 1081}

\bibitem[\protect\citeauthoryear{{Baes}, {Verstappen}, {De Looze}, {Fritz},
  {Saftly}, {Vidal P{\'e}rez}, {Stalevski}  \& {Valcke}}{{Baes}
  et~al.}{2011}]{2011ApJS..196...22B}
{Baes} M.,  {Verstappen} J.,  {De Looze} I.,  {Fritz} J.,  {Saftly} W.,  {Vidal
  P{\'e}rez} E.,  {Stalevski} M.,   {Valcke} S.,  2011, \mn@doi [\apjs]
  {10.1088/0067-0049/196/2/22}, \href
  {http://adsabs.harvard.edu/abs/2011ApJS..196...22B} {196, 22}

\bibitem[\protect\citeauthoryear{{Baes}, {Gordon}, {Lunttila}, {Bianchi},
  {Camps}, {Juvela}  \& {Kuiper}}{{Baes} et~al.}{2016}]{2016A&A...590A..55B}
{Baes} M.,  {Gordon} K.~D.,  {Lunttila} T.,  {Bianchi} S.,  {Camps} P.,
  {Juvela} M.,   {Kuiper} R.,  2016, \mn@doi [\aap]
  {10.1051/0004-6361/201528063}, \href
  {http://adsabs.harvard.edu/abs/2016A%26A...590A..55B} {590, A55}

\bibitem[\protect\citeauthoryear{{Beeston} et~al.,}{{Beeston}
  et~al.}{2018}]{2018MNRAS.479.1077B}
{Beeston} R.~A.,  et~al., 2018, \mn@doi [\mnras] {10.1093/mnras/sty1460}, \href
  {http://adsabs.harvard.edu/abs/2018MNRAS.479.1077B} {479, 1077}

\bibitem[\protect\citeauthoryear{{Bekki}}{{Bekki}}{2015}]{2015MNRAS.449.1625B}
{Bekki} K.,  2015, \mn@doi [\mnras] {10.1093/mnras/stv165}, \href
  {http://adsabs.harvard.edu/abs/2015MNRAS.449.1625B} {449, 1625}

\bibitem[\protect\citeauthoryear{{Bell}, {McIntosh}, {Katz}  \&
  {Weinberg}}{{Bell} et~al.}{2003}]{2003ApJS..149..289B}
{Bell} E.~F.,  {McIntosh} D.~H.,  {Katz} N.,   {Weinberg} M.~D.,  2003, \mn@doi
  [\apjs] {10.1086/378847}, \href
  {http://adsabs.harvard.edu/abs/2003ApJS..149..289B} {149, 289}

\bibitem[\protect\citeauthoryear{{Bianchi} et~al.,}{{Bianchi}
  et~al.}{2018}]{2018A&A...620A.112B}
{Bianchi} S.,  et~al., 2018, \mn@doi [\aap] {10.1051/0004-6361/201833699},
  \href {http://adsabs.harvard.edu/abs/2018A%26A...620A.112B} {620, A112}

\bibitem[\protect\citeauthoryear{{Blanton} et~al.,}{{Blanton}
  et~al.}{2003}]{2003ApJ...592..819B}
{Blanton} M.~R.,  et~al., 2003, \mn@doi [\apj] {10.1086/375776}, \href
  {http://adsabs.harvard.edu/abs/2003ApJ...592..819B} {592, 819}

\bibitem[\protect\citeauthoryear{{Boquien} et~al.,}{{Boquien}
  et~al.}{2016}]{2016A&A...591A...6B}
{Boquien} M.,  et~al., 2016, \mn@doi [\aap] {10.1051/0004-6361/201527759},
  \href {http://adsabs.harvard.edu/abs/2016A%26A...591A...6B} {591, A6}

\bibitem[\protect\citeauthoryear{{Boquien}, {Burgarella}, {Roehlly}, {Buat},
  {Ciesla}, {Corre}, {Inoue}  \& {Salas}}{{Boquien}
  et~al.}{2019}]{2018arXiv181103094B}
{Boquien} M.,  {Burgarella} D.,  {Roehlly} Y.,  {Buat} V.,  {Ciesla} L.,
  {Corre} D.,  {Inoue} A.~K.,   {Salas} H.,  2019, \aap, \href
  {http://adsabs.harvard.edu/abs/2018arXiv181103094B} {{in press
  (arXiv:1811.03094)}}

\bibitem[\protect\citeauthoryear{{Boselli} et~al.,}{{Boselli}
  et~al.}{2010}]{2010PASP..122..261B}
{Boselli} A.,  et~al., 2010, \mn@doi [\pasp] {10.1086/651535}, \href
  {http://adsabs.harvard.edu/abs/2010PASP..122..261B} {122, 261}

\bibitem[\protect\citeauthoryear{{Boselli} et~al.,}{{Boselli}
  et~al.}{2012}]{2012A&A...540A..54B}
{Boselli} A.,  et~al., 2012, \mn@doi [\aap] {10.1051/0004-6361/201118602},
  \href {http://adsabs.harvard.edu/abs/2012A%26A...540A..54B} {540, A54}

\bibitem[\protect\citeauthoryear{{Bourne} et~al.,}{{Bourne}
  et~al.}{2012}]{2012MNRAS.421.3027B}
{Bourne} N.,  et~al., 2012, \mn@doi [\mnras]
  {10.1111/j.1365-2966.2012.20528.x}, \href
  {http://adsabs.harvard.edu/abs/2012MNRAS.421.3027B} {421, 3027}

\bibitem[\protect\citeauthoryear{{Brinchmann}, {Charlot}, {Kauffmann},
  {Heckman}, {White}  \& {Tremonti}}{{Brinchmann}
  et~al.}{2013}]{2013MNRAS.432.2112B}
{Brinchmann} J.,  {Charlot} S.,  {Kauffmann} G.,  {Heckman} T.,  {White}
  S.~D.~M.,   {Tremonti} C.,  2013, \mn@doi [\mnras] {10.1093/mnras/stt551},
  \href {http://adsabs.harvard.edu/abs/2013MNRAS.432.2112B} {432, 2112}

\bibitem[\protect\citeauthoryear{{Brown}, {Breeveld}, {Roming}  \&
  {Siegel}}{{Brown} et~al.}{2016}]{2016AJ....152..102B}
{Brown} P.~J.,  {Breeveld} A.,  {Roming} P.~W.~A.,   {Siegel} M.,  2016,
  \mn@doi [\aj] {10.3847/0004-6256/152/4/102}, \href
  {http://adsabs.harvard.edu/abs/2016AJ....152..102B} {152, 102}

\bibitem[\protect\citeauthoryear{{Bruzual} \& {Charlot}}{{Bruzual} \&
  {Charlot}}{2003}]{2003MNRAS.344.1000B}
{Bruzual} G.,  {Charlot} S.,  2003, \mn@doi [\mnras]
  {10.1046/j.1365-8711.2003.06897.x}, \href
  {http://adsabs.harvard.edu/abs/2003MNRAS.344.1000B} {344, 1000}

\bibitem[\protect\citeauthoryear{{Buat} \& {Xu}}{{Buat} \&
  {Xu}}{1996}]{1996A&A...306...61B}
{Buat} V.,  {Xu} C.,  1996, \aap, \href
  {http://adsabs.harvard.edu/abs/1996A%26A...306...61B} {306, 61}

\bibitem[\protect\citeauthoryear{{Budav{\'a}ri} et~al.,}{{Budav{\'a}ri}
  et~al.}{2005}]{2005ApJ...619L..31B}
{Budav{\'a}ri} T.,  et~al., 2005, \mn@doi [\apjl] {10.1086/423319}, \href
  {http://adsabs.harvard.edu/abs/2005ApJ...619L..31B} {619, L31}

\bibitem[\protect\citeauthoryear{{Calura} et~al.,}{{Calura}
  et~al.}{2017}]{2017MNRAS.465...54C}
{Calura} F.,  et~al., 2017, \mn@doi [\mnras] {10.1093/mnras/stw2749}, \href
  {http://adsabs.harvard.edu/abs/2017MNRAS.465...54C} {465, 54}

\bibitem[\protect\citeauthoryear{{Calzetti}, {Armus}, {Bohlin}, {Kinney},
  {Koornneef}  \& {Storchi-Bergmann}}{{Calzetti}
  et~al.}{2000}]{2000ApJ...533..682C}
{Calzetti} D.,  {Armus} L.,  {Bohlin} R.~C.,  {Kinney} A.~L.,  {Koornneef} J.,
   {Storchi-Bergmann} T.,  2000, \mn@doi [\apj] {10.1086/308692}, \href
  {http://adsabs.harvard.edu/abs/2000ApJ...533..682C} {533, 682}

\bibitem[\protect\citeauthoryear{{Camps} \& {Baes}}{{Camps} \&
  {Baes}}{2015}]{2015A&C.....9...20C}
{Camps} P.,  {Baes} M.,  2015, \mn@doi [Astronomy and Computing]
  {10.1016/j.ascom.2014.10.004}, \href
  {http://adsabs.harvard.edu/abs/2015A%26C.....9...20C} {9, 20}

\bibitem[\protect\citeauthoryear{{Camps}, {Trayford}, {Baes}, {Theuns},
  {Schaller}  \& {Schaye}}{{Camps} et~al.}{2016}]{2016MNRAS.462.1057C}
{Camps} P.,  {Trayford} J.~W.,  {Baes} M.,  {Theuns} T.,  {Schaller} M.,
  {Schaye} J.,  2016, \mn@doi [\mnras] {10.1093/mnras/stw1735}, \href
  {http://adsabs.harvard.edu/abs/2016MNRAS.462.1057C} {462, 1057}

\bibitem[\protect\citeauthoryear{{Camps} et~al.,}{{Camps}
  et~al.}{2018}]{2018ApJS..234...20C}
{Camps} P.,  et~al., 2018, \mn@doi [\apjs] {10.3847/1538-4365/aaa24c}, \href
  {http://adsabs.harvard.edu/abs/2018ApJS..234...20C} {234, 20}

\bibitem[\protect\citeauthoryear{{Carnall}, {Leja}, {Johnson}, {McLure},
  {Dunlop}  \& {Conroy}}{{Carnall} et~al.}{2019}]{2018arXiv181103635C}
{Carnall} A.~C.,  {Leja} J.,  {Johnson} B.~D.,  {McLure} R.~J.,  {Dunlop}
  J.~S.,   {Conroy} C.,  2019, \apj, \href
  {http://adsabs.harvard.edu/abs/2018arXiv181103635C} {{submitted
  (arXiv:1811.03635)}}

\bibitem[\protect\citeauthoryear{{Chabrier}}{{Chabrier}}{2003}]{2003PASP..115..763C}
{Chabrier} G.,  2003, \mn@doi [\pasp] {10.1086/376392}, \href
  {http://adsabs.harvard.edu/abs/2003PASP..115..763C} {115, 763}

\bibitem[\protect\citeauthoryear{{Chang}, {van der Wel}, {da Cunha}  \&
  {Rix}}{{Chang} et~al.}{2015}]{2015ApJS..219....8C}
{Chang} Y.-Y.,  {van der Wel} A.,  {da Cunha} E.,   {Rix} H.-W.,  2015, \mn@doi
  [\apjs] {10.1088/0067-0049/219/1/8}, \href
  {http://adsabs.harvard.edu/abs/2015ApJS..219....8C} {219, 8}

\bibitem[\protect\citeauthoryear{{Ciesla} et~al.,}{{Ciesla}
  et~al.}{2016}]{2016A&A...585A..43C}
{Ciesla} L.,  et~al., 2016, \mn@doi [\aap] {10.1051/0004-6361/201527107}, \href
  {http://adsabs.harvard.edu/abs/2016A%26A...585A..43C} {585, A43}

\bibitem[\protect\citeauthoryear{{Clark} et~al.,}{{Clark}
  et~al.}{2015}]{2015MNRAS.452..397C}
{Clark} C.~J.~R.,  et~al., 2015, \mn@doi [\mnras] {10.1093/mnras/stv1276},
  \href {http://adsabs.harvard.edu/abs/2015MNRAS.452..397C} {452, 397}

\bibitem[\protect\citeauthoryear{{Clark} et~al.,}{{Clark}
  et~al.}{2018}]{2018A&A...609A..37C}
{Clark} C.~J.~R.,  et~al., 2018, \mn@doi [\aap] {10.1051/0004-6361/201731419},
  \href {http://adsabs.harvard.edu/abs/2018A%26A...609A..37C} {609, A37}

\bibitem[\protect\citeauthoryear{{Clemens} et~al.,}{{Clemens}
  et~al.}{2013}]{2013MNRAS.433..695C}
{Clemens} M.~S.,  et~al., 2013, \mn@doi [\mnras] {10.1093/mnras/stt760}, \href
  {http://adsabs.harvard.edu/abs/2013MNRAS.433..695C} {433, 695}

\bibitem[\protect\citeauthoryear{{Cole}, {Aragon-Salamanca}, {Frenk}, {Navarro}
   \& {Zepf}}{{Cole} et~al.}{1994}]{1994MNRAS.271..781C}
{Cole} S.,  {Aragon-Salamanca} A.,  {Frenk} C.~S.,  {Navarro} J.~F.,   {Zepf}
  S.~E.,  1994, \mn@doi [\mnras] {10.1093/mnras/271.4.781}, \href
  {http://adsabs.harvard.edu/abs/1994MNRAS.271..781C} {271, 781}

\bibitem[\protect\citeauthoryear{{Cole}, {Lacey}, {Baugh}  \& {Frenk}}{{Cole}
  et~al.}{2000}]{2000MNRAS.319..168C}
{Cole} S.,  {Lacey} C.~G.,  {Baugh} C.~M.,   {Frenk} C.~S.,  2000, \mn@doi
  [\mnras] {10.1046/j.1365-8711.2000.03879.x}, \href
  {http://adsabs.harvard.edu/abs/2000MNRAS.319..168C} {319, 168}

\bibitem[\protect\citeauthoryear{{Cole} et~al.,}{{Cole}
  et~al.}{2001}]{2001MNRAS.326..255C}
{Cole} S.,  et~al., 2001, \mn@doi [\mnras] {10.1046/j.1365-8711.2001.04591.x},
  \href {http://adsabs.harvard.edu/abs/2001MNRAS.326..255C} {326, 255}

\bibitem[\protect\citeauthoryear{{Conroy}, {Gunn}  \& {White}}{{Conroy}
  et~al.}{2009}]{2009ApJ...699..486C}
{Conroy} C.,  {Gunn} J.~E.,   {White} M.,  2009, \mn@doi [\apj]
  {10.1088/0004-637X/699/1/486}, \href
  {http://adsabs.harvard.edu/abs/2009ApJ...699..486C} {699, 486}

\bibitem[\protect\citeauthoryear{{Cortese} et~al.,}{{Cortese}
  et~al.}{2012}]{2012A&A...540A..52C}
{Cortese} L.,  et~al., 2012, \mn@doi [\aap] {10.1051/0004-6361/201118499},
  \href {http://adsabs.harvard.edu/abs/2012A%26A...540A..52C} {540, A52}

\bibitem[\protect\citeauthoryear{{Cortese} et~al.,}{{Cortese}
  et~al.}{2014}]{2014MNRAS.440..942C}
{Cortese} L.,  et~al., 2014, \mn@doi [\mnras] {10.1093/mnras/stu175}, \href
  {http://adsabs.harvard.edu/abs/2014MNRAS.440..942C} {440, 942}

\bibitem[\protect\citeauthoryear{{Cowley}, {Lacey}, {Baugh}, {Cole}, {Frenk}
  \& {Lagos}}{{Cowley} et~al.}{2019}]{2018arXiv180805208C}
{Cowley} W.~I.,  {Lacey} C.~G.,  {Baugh} C.~M.,  {Cole} S.,  {Frenk} C.~S.,
  {Lagos} C.~d.~P.,  2019, \mnras, \href
  {http://adsabs.harvard.edu/abs/2018arXiv180805208C} {{submitted}}

\bibitem[\protect\citeauthoryear{{Crain} et~al.,}{{Crain}
  et~al.}{2015}]{2015MNRAS.450.1937C}
{Crain} R.~A.,  et~al., 2015, \mn@doi [\mnras] {10.1093/mnras/stv725}, \href
  {http://adsabs.harvard.edu/abs/2015MNRAS.450.1937C} {450, 1937}

\bibitem[\protect\citeauthoryear{{Crain} et~al.,}{{Crain}
  et~al.}{2017}]{2017MNRAS.464.4204C}
{Crain} R.~A.,  et~al., 2017, \mn@doi [\mnras] {10.1093/mnras/stw2586}, \href
  {http://adsabs.harvard.edu/abs/2017MNRAS.464.4204C} {464, 4204}

\bibitem[\protect\citeauthoryear{{Cross} \& {Driver}}{{Cross} \&
  {Driver}}{2002}]{2002MNRAS.329..579C}
{Cross} N.,  {Driver} S.~P.,  2002, \mn@doi [\mnras]
  {10.1046/j.1365-8711.2002.05052.x}, \href
  {http://adsabs.harvard.edu/abs/2002MNRAS.329..579C} {329, 579}

\bibitem[\protect\citeauthoryear{{Dale}, {Helou}, {Contursi}, {Silbermann}  \&
  {Kolhatkar}}{{Dale} et~al.}{2001}]{2001ApJ...549..215D}
{Dale} D.~A.,  {Helou} G.,  {Contursi} A.,  {Silbermann} N.~A.,   {Kolhatkar}
  S.,  2001, \mn@doi [\apj] {10.1086/319077}, \href
  {http://adsabs.harvard.edu/abs/2001ApJ...549..215D} {549, 215}

\bibitem[\protect\citeauthoryear{{Dav{\'e}}, {Thompson}  \&
  {Hopkins}}{{Dav{\'e}} et~al.}{2016}]{2016MNRAS.462.3265D}
{Dav{\'e}} R.,  {Thompson} R.,   {Hopkins} P.~F.,  2016, \mn@doi [\mnras]
  {10.1093/mnras/stw1862}, \href
  {http://adsabs.harvard.edu/abs/2016MNRAS.462.3265D} {462, 3265}

\bibitem[\protect\citeauthoryear{{Davies} et~al.,}{{Davies}
  et~al.}{2012}]{2012MNRAS.419.3505D}
{Davies} J.~I.,  et~al., 2012, \mn@doi [\mnras]
  {10.1111/j.1365-2966.2011.19993.x}, \href
  {http://adsabs.harvard.edu/abs/2012MNRAS.419.3505D} {419, 3505}

\bibitem[\protect\citeauthoryear{{Davies} et~al.,}{{Davies}
  et~al.}{2015}]{2015MNRAS.447.1014D}
{Davies} L.~J.~M.,  et~al., 2015, \mn@doi [\mnras] {10.1093/mnras/stu2515},
  \href {http://adsabs.harvard.edu/abs/2015MNRAS.447.1014D} {447, 1014}

\bibitem[\protect\citeauthoryear{{Davies} et~al.,}{{Davies}
  et~al.}{2016}]{2016MNRAS.461..458D}
{Davies} L.~J.~M.,  et~al., 2016, \mn@doi [\mnras] {10.1093/mnras/stw1342},
  \href {http://adsabs.harvard.edu/abs/2016MNRAS.461..458D} {461, 458}

\bibitem[\protect\citeauthoryear{{Davies} et~al.,}{{Davies}
  et~al.}{2017}]{2017PASP..129d4102D}
{Davies} J.~I.,  et~al., 2017, \mn@doi [\pasp]
  {10.1088/1538-3873/129/974/044102}, \href
  {http://adsabs.harvard.edu/abs/2017PASP..129d4102D} {129, 044102}

\bibitem[\protect\citeauthoryear{{De Looze} et~al.,}{{De Looze}
  et~al.}{2014}]{2014A&A...571A..69D}
{De Looze} I.,  et~al., 2014, \mn@doi [\aap] {10.1051/0004-6361/201424747},
  \href {http://adsabs.harvard.edu/abs/2014A%26A...571A..69D} {571, A69}

\bibitem[\protect\citeauthoryear{{De Vis} et~al.,}{{De Vis}
  et~al.}{2017}]{2017MNRAS.471.1743D}
{De Vis} P.,  et~al., 2017, \mn@doi [\mnras] {10.1093/mnras/stx981}, \href
  {http://adsabs.harvard.edu/abs/2017MNRAS.471.1743D} {471, 1743}

\bibitem[\protect\citeauthoryear{{De Vis} et~al.,}{{De Vis}
  et~al.}{2019}]{DeVis2019}
{De Vis} P.,  et~al., 2019, \aap, {submitted}

\bibitem[\protect\citeauthoryear{{Dom{\'{\i}}nguez-Tenreiro}, {Obreja},
  {Granato}, {Schurer}, {Alpresa}, {Silva}, {Brook}  \&
  {Serna}}{{Dom{\'{\i}}nguez-Tenreiro} et~al.}{2014}]{2014MNRAS.439.3868D}
{Dom{\'{\i}}nguez-Tenreiro} R.,  {Obreja} A.,  {Granato} G.~L.,  {Schurer} A.,
  {Alpresa} P.,  {Silva} L.,  {Brook} C.~B.,   {Serna} A.,  2014, \mn@doi
  [\mnras] {10.1093/mnras/stu240}, \href
  {http://adsabs.harvard.edu/abs/2014MNRAS.439.3868D} {439, 3868}

\bibitem[\protect\citeauthoryear{{Dom{\'{\i}}nguez} et~al.,}{{Dom{\'{\i}}nguez}
  et~al.}{2011}]{2011MNRAS.410.2556D}
{Dom{\'{\i}}nguez} A.,  et~al., 2011, \mn@doi [\mnras]
  {10.1111/j.1365-2966.2010.17631.x}, \href
  {http://adsabs.harvard.edu/abs/2011MNRAS.410.2556D} {410, 2556}

\bibitem[\protect\citeauthoryear{{Driver} \& {Robotham}}{{Driver} \&
  {Robotham}}{2010}]{2010MNRAS.407.2131D}
{Driver} S.~P.,  {Robotham} A.~S.~G.,  2010, \mn@doi [\mnras]
  {10.1111/j.1365-2966.2010.17028.x}, \href
  {http://adsabs.harvard.edu/abs/2010MNRAS.407.2131D} {407, 2131}

\bibitem[\protect\citeauthoryear{{Driver} et~al.,}{{Driver}
  et~al.}{2011}]{2011MNRAS.413..971D}
{Driver} S.~P.,  et~al., 2011, \mn@doi [\mnras]
  {10.1111/j.1365-2966.2010.18188.x}, \href
  {http://adsabs.harvard.edu/abs/2011MNRAS.413..971D} {413, 971}

\bibitem[\protect\citeauthoryear{{Driver} et~al.,}{{Driver}
  et~al.}{2012}]{2012MNRAS.427.3244D}
{Driver} S.~P.,  et~al., 2012, \mn@doi [\mnras]
  {10.1111/j.1365-2966.2012.22036.x}, \href
  {http://adsabs.harvard.edu/abs/2012MNRAS.427.3244D} {427, 3244}

\bibitem[\protect\citeauthoryear{{Driver} et~al.,}{{Driver}
  et~al.}{2016}]{2016MNRAS.455.3911D}
{Driver} S.~P.,  et~al., 2016, \mn@doi [\mnras] {10.1093/mnras/stv2505}, \href
  {http://adsabs.harvard.edu/abs/2016MNRAS.455.3911D} {455, 3911}

\bibitem[\protect\citeauthoryear{{Driver} et~al.,}{{Driver}
  et~al.}{2018}]{2018MNRAS.475.2891D}
{Driver} S.~P.,  et~al., 2018, \mn@doi [\mnras] {10.1093/mnras/stx2728}, \href
  {http://adsabs.harvard.edu/abs/2018MNRAS.475.2891D} {475, 2891}

\bibitem[\protect\citeauthoryear{{Dunne} et~al.,}{{Dunne}
  et~al.}{2011}]{2011MNRAS.417.1510D}
{Dunne} L.,  et~al., 2011, \mn@doi [\mnras] {10.1111/j.1365-2966.2011.19363.x},
  \href {http://adsabs.harvard.edu/abs/2011MNRAS.417.1510D} {417, 1510}

\bibitem[\protect\citeauthoryear{{Dye} et~al.,}{{Dye}
  et~al.}{2010}]{2010A&A...518L..10D}
{Dye} S.,  et~al., 2010, \mn@doi [\aap] {10.1051/0004-6361/201014614}, \href
  {http://adsabs.harvard.edu/abs/2010A%26A...518L..10D} {518, L10}

\bibitem[\protect\citeauthoryear{{Eales} et~al.,}{{Eales}
  et~al.}{2009}]{2009ApJ...707.1779E}
{Eales} S.,  et~al., 2009, \mn@doi [\apj] {10.1088/0004-637X/707/2/1779}, \href
  {http://adsabs.harvard.edu/abs/2009ApJ...707.1779E} {707, 1779}

\bibitem[\protect\citeauthoryear{{Eales} et~al.,}{{Eales}
  et~al.}{2010}]{2010A&A...518L..23E}
{Eales} S.~A.,  et~al., 2010, \mn@doi [\aap] {10.1051/0004-6361/201014675},
  \href {http://adsabs.harvard.edu/abs/2010A%26A...518L..23E} {518, L23}

\bibitem[\protect\citeauthoryear{{Eke}, {Baugh}, {Cole}, {Frenk}, {King}  \&
  {Peacock}}{{Eke} et~al.}{2005}]{2005MNRAS.362.1233E}
{Eke} V.~R.,  {Baugh} C.~M.,  {Cole} S.,  {Frenk} C.~S.,  {King} H.~M.,
  {Peacock} J.~A.,  2005, \mn@doi [\mnras] {10.1111/j.1365-2966.2005.09384.x},
  \href {http://adsabs.harvard.edu/abs/2005MNRAS.362.1233E} {362, 1233}

\bibitem[\protect\citeauthoryear{{Ferland}, {Korista}, {Verner}, {Ferguson},
  {Kingdon}  \& {Verner}}{{Ferland} et~al.}{1998}]{1998PASP..110..761F}
{Ferland} G.~J.,  {Korista} K.~T.,  {Verner} D.~A.,  {Ferguson} J.~W.,
  {Kingdon} J.~B.,   {Verner} E.~M.,  1998, \mn@doi [\pasp] {10.1086/316190},
  \href {http://adsabs.harvard.edu/abs/1998PASP..110..761F} {110, 761}

\bibitem[\protect\citeauthoryear{{Furlong} et~al.,}{{Furlong}
  et~al.}{2015}]{2015MNRAS.450.4486F}
{Furlong} M.,  et~al., 2015, \mn@doi [\mnras] {10.1093/mnras/stv852}, \href
  {http://adsabs.harvard.edu/abs/2015MNRAS.450.4486F} {450, 4486}

\bibitem[\protect\citeauthoryear{{Furlong} et~al.,}{{Furlong}
  et~al.}{2017}]{2017MNRAS.465..722F}
{Furlong} M.,  et~al., 2017, \mn@doi [\mnras] {10.1093/mnras/stw2740}, \href
  {http://adsabs.harvard.edu/abs/2017MNRAS.465..722F} {465, 722}

\bibitem[\protect\citeauthoryear{{Galliano} et~al.,}{{Galliano}
  et~al.}{2011}]{2011A&A...536A..88G}
{Galliano} F.,  et~al., 2011, \mn@doi [\aap] {10.1051/0004-6361/201117952},
  \href {http://adsabs.harvard.edu/abs/2011A%26A...536A..88G} {536, A88}

\bibitem[\protect\citeauthoryear{{Gonzalez-Perez}, {Lacey}, {Baugh}, {Lagos},
  {Helly}, {Campbell}  \& {Mitchell}}{{Gonzalez-Perez}
  et~al.}{2014}]{2014MNRAS.439..264G}
{Gonzalez-Perez} V.,  {Lacey} C.~G.,  {Baugh} C.~M.,  {Lagos} C.~D.~P.,
  {Helly} J.,  {Campbell} D.~J.~R.,   {Mitchell} P.~D.,  2014, \mn@doi [\mnras]
  {10.1093/mnras/stt2410}, \href
  {http://adsabs.harvard.edu/abs/2014MNRAS.439..264G} {439, 264}

\bibitem[\protect\citeauthoryear{{Goz}, {Monaco}, {Granato}, {Murante},
  {Dom{\'{\i}}nguez-Tenreiro}, {Obreja}, {Annunziatella}  \& {Tescari}}{{Goz}
  et~al.}{2017}]{2017MNRAS.469.3775G}
{Goz} D.,  {Monaco} P.,  {Granato} G.~L.,  {Murante} G.,
  {Dom{\'{\i}}nguez-Tenreiro} R.,  {Obreja} A.,  {Annunziatella} M.,
  {Tescari} E.,  2017, \mn@doi [\mnras] {10.1093/mnras/stx869}, \href
  {http://adsabs.harvard.edu/abs/2017MNRAS.469.3775G} {469, 3775}

\bibitem[\protect\citeauthoryear{{Grossi} et~al.,}{{Grossi}
  et~al.}{2010}]{2010A&A...518L..52G}
{Grossi} M.,  et~al., 2010, \mn@doi [\aap] {10.1051/0004-6361/201014653}, \href
  {http://adsabs.harvard.edu/abs/2010A%26A...518L..52G} {518, L52}

\bibitem[\protect\citeauthoryear{{Grossi} et~al.,}{{Grossi}
  et~al.}{2015}]{2015A&A...574A.126G}
{Grossi} M.,  et~al., 2015, \mn@doi [\aap] {10.1051/0004-6361/201424866}, \href
  {http://adsabs.harvard.edu/abs/2015A%26A...574A.126G} {574, A126}

\bibitem[\protect\citeauthoryear{{Groves}, {Dopita}, {Sutherland}, {Kewley},
  {Fischera}, {Leitherer}, {Brandl}  \& {van Breugel}}{{Groves}
  et~al.}{2008}]{2008ApJS..176..438G}
{Groves} B.,  {Dopita} M.~A.,  {Sutherland} R.~S.,  {Kewley} L.~J.,  {Fischera}
  J.,  {Leitherer} C.,  {Brandl} B.,   {van Breugel} W.,  2008, \mn@doi [\apjs]
  {10.1086/528711}, \href {http://adsabs.harvard.edu/abs/2008ApJS..176..438G}
  {176, 438}

\bibitem[\protect\citeauthoryear{{Guidi}, {Scannapieco}  \& {Walcher}}{{Guidi}
  et~al.}{2015}]{2015MNRAS.454.2381G}
{Guidi} G.,  {Scannapieco} C.,   {Walcher} C.~J.,  2015, \mn@doi [\mnras]
  {10.1093/mnras/stv2050}, \href
  {http://adsabs.harvard.edu/abs/2015MNRAS.454.2381G} {454, 2381}

\bibitem[\protect\citeauthoryear{{Gunawardhana} et~al.,}{{Gunawardhana}
  et~al.}{2013}]{2013MNRAS.433.2764G}
{Gunawardhana} M.~L.~P.,  et~al., 2013, \mn@doi [\mnras]
  {10.1093/mnras/stt890}, \href
  {http://adsabs.harvard.edu/abs/2013MNRAS.433.2764G} {433, 2764}

\bibitem[\protect\citeauthoryear{{Hayward}, {Kere{\v s}}, {Jonsson},
  {Narayanan}, {Cox}  \& {Hernquist}}{{Hayward}
  et~al.}{2011}]{2011ApJ...743..159H}
{Hayward} C.~C.,  {Kere{\v s}} D.,  {Jonsson} P.,  {Narayanan} D.,  {Cox}
  T.~J.,   {Hernquist} L.,  2011, \mn@doi [\apj] {10.1088/0004-637X/743/2/159},
  \href {http://adsabs.harvard.edu/abs/2011ApJ...743..159H} {743, 159}

\bibitem[\protect\citeauthoryear{{Hunt} et~al.,}{{Hunt}
  et~al.}{2019}]{2018arXiv180904088H}
{Hunt} L.~K.,  et~al., 2019, \aap, \href
  {http://adsabs.harvard.edu/abs/2018arXiv180904088H} {{in press
  (arXiv:1809.04088)}}

\bibitem[\protect\citeauthoryear{{Indebetouw}, {Whitney}, {Johnson}  \&
  {Wood}}{{Indebetouw} et~al.}{2006}]{2006ApJ...636..362I}
{Indebetouw} R.,  {Whitney} B.~A.,  {Johnson} K.~E.,   {Wood} K.,  2006,
  \mn@doi [\apj] {10.1086/497886}, \href
  {http://adsabs.harvard.edu/abs/2006ApJ...636..362I} {636, 362}

\bibitem[\protect\citeauthoryear{{Inoue}}{{Inoue}}{2005}]{2005MNRAS.359..171I}
{Inoue} A.~K.,  2005, \mn@doi [\mnras] {10.1111/j.1365-2966.2005.08890.x},
  \href {http://adsabs.harvard.edu/abs/2005MNRAS.359..171I} {359, 171}

\bibitem[\protect\citeauthoryear{{Inoue}}{{Inoue}}{2011}]{2011MNRAS.415.2920I}
{Inoue} A.~K.,  2011, \mn@doi [\mnras] {10.1111/j.1365-2966.2011.18906.x},
  \href {http://adsabs.harvard.edu/abs/2011MNRAS.415.2920I} {415, 2920}

\bibitem[\protect\citeauthoryear{{Jones}, {K{\"o}hler}, {Ysard}, {Bocchio}  \&
  {Verstraete}}{{Jones} et~al.}{2017}]{2017A&A...602A..46J}
{Jones} A.~P.,  {K{\"o}hler} M.,  {Ysard} N.,  {Bocchio} M.,   {Verstraete} L.,
   2017, \mn@doi [\aap] {10.1051/0004-6361/201630225}, \href
  {http://adsabs.harvard.edu/abs/2017A%26A...602A..46J} {602, A46}

\bibitem[\protect\citeauthoryear{{Jonsson}, {Groves}  \& {Cox}}{{Jonsson}
  et~al.}{2010}]{2010MNRAS.403...17J}
{Jonsson} P.,  {Groves} B.~A.,   {Cox} T.~J.,  2010, \mn@doi [\mnras]
  {10.1111/j.1365-2966.2009.16087.x}, \href
  {http://adsabs.harvard.edu/abs/2010MNRAS.403...17J} {403, 17}

\bibitem[\protect\citeauthoryear{{Katsianis} et~al.,}{{Katsianis}
  et~al.}{2017}]{2017MNRAS.472..919K}
{Katsianis} A.,  et~al., 2017, \mn@doi [\mnras] {10.1093/mnras/stx2020}, \href
  {http://adsabs.harvard.edu/abs/2017MNRAS.472..919K} {472, 919}

\bibitem[\protect\citeauthoryear{{Kauffmann}, {White}  \&
  {Guiderdoni}}{{Kauffmann} et~al.}{1993}]{1993MNRAS.264..201K}
{Kauffmann} G.,  {White} S.~D.~M.,   {Guiderdoni} B.,  1993, \mn@doi [\mnras]
  {10.1093/mnras/264.1.201}, \href
  {http://adsabs.harvard.edu/abs/1993MNRAS.264..201K} {264, 201}

\bibitem[\protect\citeauthoryear{{Kelvin} et~al.,}{{Kelvin}
  et~al.}{2014}]{2014MNRAS.439.1245K}
{Kelvin} L.~S.,  et~al., 2014, \mn@doi [\mnras] {10.1093/mnras/stt2391}, \href
  {http://adsabs.harvard.edu/abs/2014MNRAS.439.1245K} {439, 1245}

\bibitem[\protect\citeauthoryear{{Kennicutt}}{{Kennicutt}}{1998}]{1998ApJ...498..541K}
{Kennicutt} Jr. R.~C.,  1998, \mn@doi [\apj] {10.1086/305588}, \href
  {http://adsabs.harvard.edu/abs/1998ApJ...498..541K} {498, 541}

\bibitem[\protect\citeauthoryear{{Khandai}, {Di Matteo}, {Croft}, {Wilkins},
  {Feng}, {Tucker}, {DeGraf}  \& {Liu}}{{Khandai}
  et~al.}{2015}]{2015MNRAS.450.1349K}
{Khandai} N.,  {Di Matteo} T.,  {Croft} R.,  {Wilkins} S.,  {Feng} Y.,
  {Tucker} E.,  {DeGraf} C.,   {Liu} M.-S.,  2015, \mn@doi [\mnras]
  {10.1093/mnras/stv627}, \href
  {http://adsabs.harvard.edu/abs/2015MNRAS.450.1349K} {450, 1349}

\bibitem[\protect\citeauthoryear{{Kochanek} et~al.,}{{Kochanek}
  et~al.}{2001}]{2001ApJ...560..566K}
{Kochanek} C.~S.,  et~al., 2001, \mn@doi [\apj] {10.1086/322488}, \href
  {http://adsabs.harvard.edu/abs/2001ApJ...560..566K} {560, 566}

\bibitem[\protect\citeauthoryear{{Lacey} et~al.,}{{Lacey}
  et~al.}{2016}]{2016MNRAS.462.3854L}
{Lacey} C.~G.,  et~al., 2016, \mn@doi [\mnras] {10.1093/mnras/stw1888}, \href
  {http://adsabs.harvard.edu/abs/2016MNRAS.462.3854L} {462, 3854}

\bibitem[\protect\citeauthoryear{{Lagos} et~al.,}{{Lagos}
  et~al.}{2015}]{2015MNRAS.452.3815L}
{Lagos} C.~d.~P.,  et~al., 2015, \mn@doi [\mnras] {10.1093/mnras/stv1488},
  \href {http://adsabs.harvard.edu/abs/2015MNRAS.452.3815L} {452, 3815}

\bibitem[\protect\citeauthoryear{{Lagos}, {Theuns}, {Stevens}, {Cortese},
  {Padilla}, {Davis}, {Contreras}  \& {Croton}}{{Lagos}
  et~al.}{2017}]{2017MNRAS.464.3850L}
{Lagos} C.~d.~P.,  {Theuns} T.,  {Stevens} A.~R.~H.,  {Cortese} L.,  {Padilla}
  N.~D.,  {Davis} T.~A.,  {Contreras} S.,   {Croton} D.,  2017, \mn@doi
  [\mnras] {10.1093/mnras/stw2610}, \href
  {http://adsabs.harvard.edu/abs/2017MNRAS.464.3850L} {464, 3850}

\bibitem[\protect\citeauthoryear{{Leja}, {Carnall}, {Johnson}, {Conroy}  \&
  {Speagle}}{{Leja} et~al.}{2019}]{2018arXiv181103637L}
{Leja} J.,  {Carnall} A.~C.,  {Johnson} B.~D.,  {Conroy} C.,   {Speagle} J.~S.,
   2019, \apj, \href {http://adsabs.harvard.edu/abs/2018arXiv181103637L}
  {{submitted (arXiv:1811.03637)}}

\bibitem[\protect\citeauthoryear{{Li} \& {White}}{{Li} \&
  {White}}{2009}]{2009MNRAS.398.2177L}
{Li} C.,  {White} S.~D.~M.,  2009, \mn@doi [\mnras]
  {10.1111/j.1365-2966.2009.15268.x}, \href
  {http://adsabs.harvard.edu/abs/2009MNRAS.398.2177L} {398, 2177}

\bibitem[\protect\citeauthoryear{{Liske} et~al.,}{{Liske}
  et~al.}{2015}]{2015MNRAS.452.2087L}
{Liske} J.,  et~al., 2015, \mn@doi [\mnras] {10.1093/mnras/stv1436}, \href
  {http://adsabs.harvard.edu/abs/2015MNRAS.452.2087L} {452, 2087}

\bibitem[\protect\citeauthoryear{{Madau} \& {Dickinson}}{{Madau} \&
  {Dickinson}}{2014}]{2014ARA&A..52..415M}
{Madau} P.,  {Dickinson} M.,  2014, \mn@doi [\araa]
  {10.1146/annurev-astro-081811-125615}, \href
  {http://adsabs.harvard.edu/abs/2014ARA%26A..52..415M} {52, 415}

\bibitem[\protect\citeauthoryear{{Marchetti} et~al.,}{{Marchetti}
  et~al.}{2016}]{2016MNRAS.456.1999M}
{Marchetti} L.,  et~al., 2016, \mn@doi [\mnras] {10.1093/mnras/stv2717}, \href
  {http://adsabs.harvard.edu/abs/2016MNRAS.456.1999M} {456, 1999}

\bibitem[\protect\citeauthoryear{{McAlpine} et~al.,}{{McAlpine}
  et~al.}{2016}]{2016A&C....15...72M}
{McAlpine} S.,  et~al., 2016, \mn@doi [Astronomy and Computing]
  {10.1016/j.ascom.2016.02.004}, \href
  {http://adsabs.harvard.edu/abs/2016A%26C....15...72M} {15, 72}

\bibitem[\protect\citeauthoryear{{McKinnon}, {Torrey}  \&
  {Vogelsberger}}{{McKinnon} et~al.}{2016}]{2016MNRAS.457.3775M}
{McKinnon} R.,  {Torrey} P.,   {Vogelsberger} M.,  2016, \mn@doi [\mnras]
  {10.1093/mnras/stw253}, \href
  {http://adsabs.harvard.edu/abs/2016MNRAS.457.3775M} {457, 3775}

\bibitem[\protect\citeauthoryear{{McKinnon}, {Torrey}, {Vogelsberger},
  {Hayward}  \& {Marinacci}}{{McKinnon} et~al.}{2017}]{2017MNRAS.468.1505M}
{McKinnon} R.,  {Torrey} P.,  {Vogelsberger} M.,  {Hayward} C.~C.,
  {Marinacci} F.,  2017, \mn@doi [\mnras] {10.1093/mnras/stx467}, \href
  {http://adsabs.harvard.edu/abs/2017MNRAS.468.1505M} {468, 1505}

\bibitem[\protect\citeauthoryear{{McKinnon}, {Vogelsberger}, {Torrey},
  {Marinacci}  \& {Kannan}}{{McKinnon} et~al.}{2018}]{2018MNRAS.478.2851M}
{McKinnon} R.,  {Vogelsberger} M.,  {Torrey} P.,  {Marinacci} F.,   {Kannan}
  R.,  2018, \mn@doi [\mnras] {10.1093/mnras/sty1248}, \href
  {http://adsabs.harvard.edu/abs/2018MNRAS.478.2851M} {478, 2851}

\bibitem[\protect\citeauthoryear{{Moffett} et~al.,}{{Moffett}
  et~al.}{2016}]{2016MNRAS.462.4336M}
{Moffett} A.~J.,  et~al., 2016, \mn@doi [\mnras] {10.1093/mnras/stw1861}, \href
  {http://adsabs.harvard.edu/abs/2016MNRAS.462.4336M} {462, 4336}

\bibitem[\protect\citeauthoryear{{Montero-Dorta} \& {Prada}}{{Montero-Dorta} \&
  {Prada}}{2009}]{2009MNRAS.399.1106M}
{Montero-Dorta} A.~D.,  {Prada} F.,  2009, \mn@doi [\mnras]
  {10.1111/j.1365-2966.2009.15197.x}, \href
  {http://adsabs.harvard.edu/abs/2009MNRAS.399.1106M} {399, 1106}

\bibitem[\protect\citeauthoryear{{Moustakas} et~al.,}{{Moustakas}
  et~al.}{2013}]{2013ApJ...767...50M}
{Moustakas} J.,  et~al., 2013, \mn@doi [\apj] {10.1088/0004-637X/767/1/50},
  \href {http://adsabs.harvard.edu/abs/2013ApJ...767...50M} {767, 50}

\bibitem[\protect\citeauthoryear{{Natale}, {Popescu}, {Tuffs}, {Debattista},
  {Fischera}  \& {Grootes}}{{Natale} et~al.}{2015}]{2015MNRAS.449..243N}
{Natale} G.,  {Popescu} C.~C.,  {Tuffs} R.~J.,  {Debattista} V.~P.,  {Fischera}
  J.,   {Grootes} M.~W.,  2015, \mn@doi [\mnras] {10.1093/mnras/stv286}, \href
  {http://adsabs.harvard.edu/abs/2015MNRAS.449..243N} {449, 243}

\bibitem[\protect\citeauthoryear{{Nersesian} et~al.,}{{Nersesian}
  et~al.}{2019}]{Nersesian2019}
{Nersesian} A.,  et~al., 2019, \aap, {submitted}

\bibitem[\protect\citeauthoryear{{Noll}, {Burgarella}, {Giovannoli}, {Buat},
  {Marcillac}  \& {Mu{\~n}oz-Mateos}}{{Noll}
  et~al.}{2009}]{2009A&A...507.1793N}
{Noll} S.,  {Burgarella} D.,  {Giovannoli} E.,  {Buat} V.,  {Marcillac} D.,
  {Mu{\~n}oz-Mateos} J.~C.,  2009, \mn@doi [\aap]
  {10.1051/0004-6361/200912497}, \href
  {http://adsabs.harvard.edu/abs/2009A%26A...507.1793N} {507, 1793}

\bibitem[\protect\citeauthoryear{{Norberg} et~al.,}{{Norberg}
  et~al.}{2002}]{2002MNRAS.336..907N}
{Norberg} P.,  et~al., 2002, \mn@doi [\mnras]
  {10.1046/j.1365-8711.2002.05831.x}, \href
  {http://adsabs.harvard.edu/abs/2002MNRAS.336..907N} {336, 907}

\bibitem[\protect\citeauthoryear{{Oliver} et~al.,}{{Oliver}
  et~al.}{2012}]{2012MNRAS.424.1614O}
{Oliver} S.~J.,  et~al., 2012, \mn@doi [\mnras]
  {10.1111/j.1365-2966.2012.20912.x}, \href
  {http://adsabs.harvard.edu/abs/2012MNRAS.424.1614O} {424, 1614}

\bibitem[\protect\citeauthoryear{{Panter}, {Heavens}  \& {Jimenez}}{{Panter}
  et~al.}{2004}]{2004MNRAS.355..764P}
{Panter} B.,  {Heavens} A.~F.,   {Jimenez} R.,  2004, \mn@doi [\mnras]
  {10.1111/j.1365-2966.2004.08355.x}, \href
  {http://adsabs.harvard.edu/abs/2004MNRAS.355..764P} {355, 764}

\bibitem[\protect\citeauthoryear{{Pappalardo} et~al.,}{{Pappalardo}
  et~al.}{2016}]{2016A&A...589A..11P}
{Pappalardo} C.,  et~al., 2016, \mn@doi [\aap] {10.1051/0004-6361/201528008},
  \href {http://adsabs.harvard.edu/abs/2016A%26A...589A..11P} {589, A11}

\bibitem[\protect\citeauthoryear{{Pforr}, {Maraston}  \& {Tonini}}{{Pforr}
  et~al.}{2012}]{2012MNRAS.422.3285P}
{Pforr} J.,  {Maraston} C.,   {Tonini} C.,  2012, \mn@doi [\mnras]
  {10.1111/j.1365-2966.2012.20848.x}, \href
  {http://adsabs.harvard.edu/abs/2012MNRAS.422.3285P} {422, 3285}

\bibitem[\protect\citeauthoryear{{Pillepich} et~al.,}{{Pillepich}
  et~al.}{2018}]{2018MNRAS.473.4077P}
{Pillepich} A.,  et~al., 2018, \mn@doi [\mnras] {10.1093/mnras/stx2656}, \href
  {http://adsabs.harvard.edu/abs/2018MNRAS.473.4077P} {473, 4077}

\bibitem[\protect\citeauthoryear{{Planck Collaboration XVI}}{{Planck
  Collaboration XVI}}{2014}]{2014A&A...571A..16P}
{Planck Collaboration XVI} 2014, \mn@doi [\aap] {10.1051/0004-6361/201321591},
  \href {http://adsabs.harvard.edu/abs/2014A%26A...571A..16P} {571, A16}

\bibitem[\protect\citeauthoryear{{Popescu} \& {Tuffs}}{{Popescu} \&
  {Tuffs}}{2002}]{2002MNRAS.335L..41P}
{Popescu} C.~C.,  {Tuffs} R.~J.,  2002, \mn@doi [\mnras]
  {10.1046/j.1365-8711.2002.05881.x}, \href
  {http://adsabs.harvard.edu/abs/2002MNRAS.335L..41P} {335, L41}

\bibitem[\protect\citeauthoryear{{Popping}, {Somerville}  \&
  {Galametz}}{{Popping} et~al.}{2017}]{2017MNRAS.471.3152P}
{Popping} G.,  {Somerville} R.~S.,   {Galametz} M.,  2017, \mn@doi [\mnras]
  {10.1093/mnras/stx1545}, \href
  {http://adsabs.harvard.edu/abs/2017MNRAS.471.3152P} {471, 3152}

\bibitem[\protect\citeauthoryear{{Robotham} \& {Driver}}{{Robotham} \&
  {Driver}}{2011}]{2011MNRAS.413.2570R}
{Robotham} A.~S.~G.,  {Driver} S.~P.,  2011, \mn@doi [\mnras]
  {10.1111/j.1365-2966.2011.18327.x}, \href
  {http://adsabs.harvard.edu/abs/2011MNRAS.413.2570R} {413, 2570}

\bibitem[\protect\citeauthoryear{{Rodriguez-Gomez} et~al.,}{{Rodriguez-Gomez}
  et~al.}{2019}]{2018arXiv180908239R}
{Rodriguez-Gomez} V.,  et~al., 2019, \mnras, \href
  {http://adsabs.harvard.edu/abs/2018arXiv180908239R} {{in press
  (arXiv:1809.08239)}}

\bibitem[\protect\citeauthoryear{{Rosas-Guevara}, {Bower}, {Schaye},
  {McAlpine}, {Dalla Vecchia}, {Frenk}, {Schaller}  \&
  {Theuns}}{{Rosas-Guevara} et~al.}{2016}]{2016MNRAS.462..190R}
{Rosas-Guevara} Y.,  {Bower} R.~G.,  {Schaye} J.,  {McAlpine} S.,  {Dalla
  Vecchia} C.,  {Frenk} C.~S.,  {Schaller} M.,   {Theuns} T.,  2016, \mn@doi
  [\mnras] {10.1093/mnras/stw1679}, \href
  {http://adsabs.harvard.edu/abs/2016MNRAS.462..190R} {462, 190}

\bibitem[\protect\citeauthoryear{{Saftly}, {Camps}, {Baes}, {Gordon},
  {Vandewoude}, {Rahimi}  \& {Stalevski}}{{Saftly}
  et~al.}{2013}]{2013A&A...554A..10S}
{Saftly} W.,  {Camps} P.,  {Baes} M.,  {Gordon} K.~D.,  {Vandewoude} S.,
  {Rahimi} A.,   {Stalevski} M.,  2013, \mn@doi [\aap]
  {10.1051/0004-6361/201220854}, \href
  {http://adsabs.harvard.edu/abs/2013A%26A...554A..10S} {554, A10}

\bibitem[\protect\citeauthoryear{{Saftly}, {Baes}  \& {Camps}}{{Saftly}
  et~al.}{2014}]{2014A&A...561A..77S}
{Saftly} W.,  {Baes} M.,   {Camps} P.,  2014, \mn@doi [\aap]
  {10.1051/0004-6361/201322593}, \href
  {http://adsabs.harvard.edu/abs/2014A%26A...561A..77S} {561, A77}

\bibitem[\protect\citeauthoryear{{Saftly}, {Baes}, {De Geyter}, {Camps},
  {Renaud}, {Guedes}  \& {De Looze}}{{Saftly}
  et~al.}{2015}]{2015A&A...576A..31S}
{Saftly} W.,  {Baes} M.,  {De Geyter} G.,  {Camps} P.,  {Renaud} F.,  {Guedes}
  J.,   {De Looze} I.,  2015, \mn@doi [\aap] {10.1051/0004-6361/201425445},
  \href {http://adsabs.harvard.edu/abs/2015A%26A...576A..31S} {576, A31}

\bibitem[\protect\citeauthoryear{{Salpeter}}{{Salpeter}}{1955}]{1955ApJ...121..161S}
{Salpeter} E.~E.,  1955, \mn@doi [\apj] {10.1086/145971}, \href
  {http://adsabs.harvard.edu/abs/1955ApJ...121..161S} {121, 161}

\bibitem[\protect\citeauthoryear{{Santini} et~al.,}{{Santini}
  et~al.}{2014}]{2014A&A...562A..30S}
{Santini} P.,  et~al., 2014, \mn@doi [\aap] {10.1051/0004-6361/201322835},
  \href {http://adsabs.harvard.edu/abs/2014A%26A...562A..30S} {562, A30}

\bibitem[\protect\citeauthoryear{{Saunders}, {Rowan-Robinson}, {Lawrence},
  {Efstathiou}, {Kaiser}, {Ellis}  \& {Frenk}}{{Saunders}
  et~al.}{1990}]{1990MNRAS.242..318S}
{Saunders} W.,  {Rowan-Robinson} M.,  {Lawrence} A.,  {Efstathiou} G.,
  {Kaiser} N.,  {Ellis} R.~S.,   {Frenk} C.~S.,  1990, \mn@doi [\mnras]
  {10.1093/mnras/242.3.318}, \href
  {http://adsabs.harvard.edu/abs/1990MNRAS.242..318S} {242, 318}

\bibitem[\protect\citeauthoryear{{Schaye}}{{Schaye}}{2004}]{2004ApJ...609..667S}
{Schaye} J.,  2004, \mn@doi [\apj] {10.1086/421232}, \href
  {http://adsabs.harvard.edu/abs/2004ApJ...609..667S} {609, 667}

\bibitem[\protect\citeauthoryear{{Schaye} \& {Dalla Vecchia}}{{Schaye} \&
  {Dalla Vecchia}}{2008}]{2008MNRAS.383.1210S}
{Schaye} J.,  {Dalla Vecchia} C.,  2008, \mn@doi [\mnras]
  {10.1111/j.1365-2966.2007.12639.x}, \href
  {http://adsabs.harvard.edu/abs/2008MNRAS.383.1210S} {383, 1210}

\bibitem[\protect\citeauthoryear{{Schaye} et~al.,}{{Schaye}
  et~al.}{2015}]{2015MNRAS.446..521S}
{Schaye} J.,  et~al., 2015, \mn@doi [\mnras] {10.1093/mnras/stu2058}, \href
  {http://adsabs.harvard.edu/abs/2015MNRAS.446..521S} {446, 521}

\bibitem[\protect\citeauthoryear{{Schmidt}}{{Schmidt}}{1959}]{1959ApJ...129..243S}
{Schmidt} M.,  1959, \mn@doi [\apj] {10.1086/146614}, \href
  {http://adsabs.harvard.edu/abs/1959ApJ...129..243S} {129, 243}

\bibitem[\protect\citeauthoryear{{Scoville} et~al.,}{{Scoville}
  et~al.}{2007}]{2007ApJS..172....1S}
{Scoville} N.,  et~al., 2007, \mn@doi [\apjs] {10.1086/516585}, \href
  {http://adsabs.harvard.edu/abs/2007ApJS..172....1S} {172, 1}

\bibitem[\protect\citeauthoryear{{Skibba} et~al.,}{{Skibba}
  et~al.}{2011}]{2011ApJ...738...89S}
{Skibba} R.~A.,  et~al., 2011, \mn@doi [\apj] {10.1088/0004-637X/738/1/89},
  \href {http://adsabs.harvard.edu/abs/2011ApJ...738...89S} {738, 89}

\bibitem[\protect\citeauthoryear{{Smith}, {Loveday}  \& {Cross}}{{Smith}
  et~al.}{2009}]{2009MNRAS.397..868S}
{Smith} A.~J.,  {Loveday} J.,   {Cross} N.~J.~G.,  2009, \mn@doi [\mnras]
  {10.1111/j.1365-2966.2009.14987.x}, \href
  {http://adsabs.harvard.edu/abs/2009MNRAS.397..868S} {397, 868}

\bibitem[\protect\citeauthoryear{{Smith} et~al.,}{{Smith}
  et~al.}{2012a}]{2012MNRAS.427..703S}
{Smith} D.~J.~B.,  et~al., 2012a, \mn@doi [\mnras]
  {10.1111/j.1365-2966.2012.21930.x}, \href
  {http://adsabs.harvard.edu/abs/2012MNRAS.427..703S} {427, 703}

\bibitem[\protect\citeauthoryear{{Smith} et~al.,}{{Smith}
  et~al.}{2012b}]{2012ApJ...748..123S}
{Smith} M.~W.~L.,  et~al., 2012b, \mn@doi [\apj] {10.1088/0004-637X/748/2/123},
  \href {http://adsabs.harvard.edu/abs/2012ApJ...748..123S} {748, 123}

\bibitem[\protect\citeauthoryear{{Soifer} \& {Neugebauer}}{{Soifer} \&
  {Neugebauer}}{1991}]{1991AJ....101..354S}
{Soifer} B.~T.,  {Neugebauer} G.,  1991, \mn@doi [\aj] {10.1086/115691}, \href
  {http://adsabs.harvard.edu/abs/1991AJ....101..354S} {101, 354}

\bibitem[\protect\citeauthoryear{{Somerville} \& {Primack}}{{Somerville} \&
  {Primack}}{1999}]{1999MNRAS.310.1087S}
{Somerville} R.~S.,  {Primack} J.~R.,  1999, \mn@doi [\mnras]
  {10.1046/j.1365-8711.1999.03032.x}, \href
  {http://adsabs.harvard.edu/abs/1999MNRAS.310.1087S} {310, 1087}

\bibitem[\protect\citeauthoryear{{Springel}}{{Springel}}{2005}]{2005MNRAS.364.1105S}
{Springel} V.,  2005, \mn@doi [\mnras] {10.1111/j.1365-2966.2005.09655.x},
  \href {http://adsabs.harvard.edu/abs/2005MNRAS.364.1105S} {364, 1105}

\bibitem[\protect\citeauthoryear{{Steinacker}, {Baes}  \&
  {Gordon}}{{Steinacker} et~al.}{2013}]{2013ARA&A..51...63S}
{Steinacker} J.,  {Baes} M.,   {Gordon} K.~D.,  2013, \mn@doi [\araa]
  {10.1146/annurev-astro-082812-141042}, \href
  {http://adsabs.harvard.edu/abs/2013ARA%26A..51...63S} {51, 63}

\bibitem[\protect\citeauthoryear{{Takeuchi}, {Ishii}, {Dole}, {Dennefeld},
  {Lagache}  \& {Puget}}{{Takeuchi} et~al.}{2006}]{2006A&A...448..525T}
{Takeuchi} T.~T.,  {Ishii} T.~T.,  {Dole} H.,  {Dennefeld} M.,  {Lagache} G.,
  {Puget} J.-L.,  2006, \mn@doi [\aap] {10.1051/0004-6361:20054272}, \href
  {http://adsabs.harvard.edu/abs/2006A%26A...448..525T} {448, 525}

\bibitem[\protect\citeauthoryear{{Trayford} et~al.,}{{Trayford}
  et~al.}{2015}]{2015MNRAS.452.2879T}
{Trayford} J.~W.,  et~al., 2015, \mn@doi [\mnras] {10.1093/mnras/stv1461},
  \href {http://adsabs.harvard.edu/abs/2015MNRAS.452.2879T} {452, 2879}

\bibitem[\protect\citeauthoryear{{Trayford} et~al.,}{{Trayford}
  et~al.}{2017}]{2017MNRAS.470..771T}
{Trayford} J.~W.,  et~al., 2017, \mn@doi [\mnras] {10.1093/mnras/stx1051},
  \href {http://adsabs.harvard.edu/abs/2017MNRAS.470..771T} {470, 771}

\bibitem[\protect\citeauthoryear{{Verstocken}, {Van De Putte}, {Camps}  \&
  {Baes}}{{Verstocken} et~al.}{2017}]{2017A&C....20...16V}
{Verstocken} S.,  {Van De Putte} D.,  {Camps} P.,   {Baes} M.,  2017, \mn@doi
  [Astronomy and Computing] {10.1016/j.ascom.2017.05.003}, \href
  {http://adsabs.harvard.edu/abs/2017A%26C....20...16V} {20, 16}

\bibitem[\protect\citeauthoryear{{Viaene} et~al.,}{{Viaene}
  et~al.}{2016}]{2016A&A...586A..13V}
{Viaene} S.,  et~al., 2016, \mn@doi [\aap] {10.1051/0004-6361/201527586}, \href
  {http://adsabs.harvard.edu/abs/2016A%26A...586A..13V} {586, A13}

\bibitem[\protect\citeauthoryear{{Vogelsberger} et~al.,}{{Vogelsberger}
  et~al.}{2014}]{2014MNRAS.444.1518V}
{Vogelsberger} M.,  et~al., 2014, \mn@doi [\mnras] {10.1093/mnras/stu1536},
  \href {http://adsabs.harvard.edu/abs/2014MNRAS.444.1518V} {444, 1518}

\bibitem[\protect\citeauthoryear{{Wiersma}, {Schaye}, {Theuns}, {Dalla Vecchia}
   \& {Tornatore}}{{Wiersma} et~al.}{2009}]{2009MNRAS.399..574W}
{Wiersma} R.~P.~C.,  {Schaye} J.,  {Theuns} T.,  {Dalla Vecchia} C.,
  {Tornatore} L.,  2009, \mn@doi [\mnras] {10.1111/j.1365-2966.2009.15331.x},
  \href {http://adsabs.harvard.edu/abs/2009MNRAS.399..574W} {399, 574}

\bibitem[\protect\citeauthoryear{{Wilson}, {Cowie}, {Barger}  \&
  {Burke}}{{Wilson} et~al.}{2002}]{2002AJ....124.1258W}
{Wilson} G.,  {Cowie} L.~L.,  {Barger} A.~J.,   {Burke} D.~J.,  2002, \mn@doi
  [\aj] {10.1086/341818}, \href
  {http://adsabs.harvard.edu/abs/2002AJ....124.1258W} {124, 1258}

\bibitem[\protect\citeauthoryear{{Witt} \& {Gordon}}{{Witt} \&
  {Gordon}}{2000}]{2000ApJ...528..799W}
{Witt} A.~N.,  {Gordon} K.~D.,  2000, \mn@doi [\apj] {10.1086/308197}, \href
  {http://adsabs.harvard.edu/abs/2000ApJ...528..799W} {528, 799}

\bibitem[\protect\citeauthoryear{{Wright} et~al.,}{{Wright}
  et~al.}{2017}]{2017MNRAS.470..283W}
{Wright} A.~H.,  et~al., 2017, \mn@doi [\mnras] {10.1093/mnras/stx1149}, \href
  {http://adsabs.harvard.edu/abs/2017MNRAS.470..283W} {470, 283}

\bibitem[\protect\citeauthoryear{{Wyder} et~al.,}{{Wyder}
  et~al.}{2005}]{2005ApJ...619L..15W}
{Wyder} T.~K.,  et~al., 2005, \mn@doi [\apjl] {10.1086/424735}, \href
  {http://adsabs.harvard.edu/abs/2005ApJ...619L..15W} {619, L15}

\bibitem[\protect\citeauthoryear{{Xu} \& {Buat}}{{Xu} \&
  {Buat}}{1995}]{1995A&A...293L..65X}
{Xu} C.,  {Buat} V.,  1995, \aap, \href
  {http://adsabs.harvard.edu/abs/1995A%26A...293L..65X} {293}

\bibitem[\protect\citeauthoryear{{Zafar} \& {Watson}}{{Zafar} \&
  {Watson}}{2013}]{2013A&A...560A..26Z}
{Zafar} T.,  {Watson} D.,  2013, \mn@doi [\aap] {10.1051/0004-6361/201321413},
  \href {http://adsabs.harvard.edu/abs/2013A%26A...560A..26Z} {560, A26}

\bibitem[\protect\citeauthoryear{{Zhukovska}, {Dobbs}, {Jenkins}  \&
  {Klessen}}{{Zhukovska} et~al.}{2016}]{2016ApJ...831..147Z}
{Zhukovska} S.,  {Dobbs} C.,  {Jenkins} E.~B.,   {Klessen} R.~S.,  2016,
  \mn@doi [\apj] {10.3847/0004-637X/831/2/147}, \href
  {http://adsabs.harvard.edu/abs/2016ApJ...831..147Z} {831, 147}

\bibitem[\protect\citeauthoryear{{da Cunha}, {Charlot}  \& {Elbaz}}{{da Cunha}
  et~al.}{2008}]{2008MNRAS.388.1595D}
{da Cunha} E.,  {Charlot} S.,   {Elbaz} D.,  2008, \mn@doi [\mnras]
  {10.1111/j.1365-2966.2008.13535.x}, \href
  {http://adsabs.harvard.edu/abs/2008MNRAS.388.1595D} {388, 1595}

\bibitem[\protect\citeauthoryear{{di Serego Alighieri} et~al.,}{{di Serego
  Alighieri} et~al.}{2013}]{2013A&A...552A...8D}
{di Serego Alighieri} S.,  et~al., 2013, \mn@doi [\aap]
  {10.1051/0004-6361/201220551}, \href
  {http://adsabs.harvard.edu/abs/2013A%26A...552A...8D} {552, A8}

\makeatother
\end{thebibliography}

\label{lastpage}
\end{document}